\documentclass[%
 reprint,
 superscriptaddress,
 amsmath,amssymb,
 aip,
 chaos,
floatfix,
]{revtex4-2}

\usepackage{color,graphics}
\usepackage{graphicx}
\usepackage[sort&compress]{natbib}
\usepackage{hhline}
\usepackage{soul}
\usepackage{url}
\usepackage{hyperref}

\usepackage{graphicx}
\usepackage{dcolumn}
\usepackage{bm}

\usepackage[utf8]{inputenc}
\usepackage[T1]{fontenc}
\usepackage{newtxtext,newtxmath} 
\usepackage{etoolbox}
\usepackage{algorithm}
\usepackage{algpseudocode}

\providecommand{\onlinecite}[1]{\hspace{-1 ex} \nocite{#1}\citenum{#1}}

\usepackage[usenames,dvipsnames]{xcolor}

\makeatletter
\def\@email#1#2#3#4#5#6#7{%
 \endgroup
 \patchcmd{\titleblock@produce}
  {\frontmatter@RRAPformat}
  {\frontmatter@RRAPformat
  {\produce@RRAP{#1
#2\href{mailto:#3}{#3}
  #4\href{mailto:#5}{#5}
  #6\href{mailto:#7}{#7}
  }}  
  \frontmatter@RRAPformat}
  {}{}
}%
\makeatother

\begin{document}

\preprint{
}

\title[]{Early-warning the compact-to-dendritic transition via spatiotemporal learning of two-dimensional growth images}

\author{Hyunjun Jang}
\affiliation{Department of Energy Engineering, Korea Institute of Energy Technology (KENTECH), Naju 58330, Republic of Korea}
\author{Chung Bin Park$^{\dagger}$}
\affiliation{Department of Chemistry Education, Korea National University of Education, Cheongju, 28173, Republic of Korea}
\author{Jeonghoon Kim$^{\ddagger}$}
\affiliation{Microsoft, Redmond, WA, 98052, USA}
\author{Jeongmin Kim$^{*}$}
\affiliation{Department of Chemistry Education and Graduate Department of Chemical Materials, Pusan National University, Busan 46241, Republic of Korea}
\email{$^\dagger$}{cbpark@knue.ac.kr} {$^\ddagger$}{jeonghoonkim@microsoft.com} {$^*$}{jeongmin@pusan.ac.kr}

\date{\today}

\begin{abstract}
Transitions between distinct dynamical regimes are ubiquitous in nonequilibrium systems. As a prototypical example, deposition growth is often accompanied by irreversible morphological instabilities. Forecasting such transitions from pre-transition configurations remains fundamentally challenging, as early precursors are weak, spatially heterogeneous, and masked by inherent fluctuations. Here, we investigate compact-to-dendritic transitions (CDTs) in a two-dimensional particle-based electrodeposition model and formulate a horizon-based early-warning task using trajectory-resolved transition points. We demonstrate that anticipating the CDT is intrinsically a spatiotemporal problem: neither static morphological descriptors nor temporal learning applied to predefined features alone yields reliable predictive signals. In contrast, end-to-end learning of jointly optimized spatial and temporal representations from growth images enables robust anticipation across a wide range of prediction horizons. Analysis of the learned latent dynamics reveals the emergence of a low-dimensional surrogate variable that tracks progressive morphological destabilization and undergoes reorganization near the transition. We further show that the learned spatiotemporal representation exhibits limited but systematic transferability across reaction-rate conditions, with predictive performance degrading as the inference condition departs from the training condition, consistent with changes in the latent-state dynamics.  Overall, our results establish a general formulation for forecasting incipient instabilities in nonequilibrium interfacial growth, with implications for the predictive monitoring and control of pattern-forming driven systems.
\end{abstract}

\maketitle

\section{Introduction}\label{sec:intro}

Sudden transitions between distinct dynamical regimes are a hallmark of nonequilibrium systems~\cite{May1977, PhysRevLett.128.110603, PhysRevLett.110.135704, PhysRevLett.98.195702, PhysRevE.110.064156, RevModPhys.76.663, PhysRevE.104.064123, doi:10.1073/pnas.0811729106}. 
Such transitions are often accompanied by qualitative changes in structure, transport, or reactivity, and arise across a wide range of contexts, including pattern-forming instabilities, percolation and fracture processes, and driven interfacial growth~\cite{PhysRevLett.118.013902, PhysRevLett.129.190601, PhysRevLett.83.4999, PhysRevE.59.5049, PhysRevLett.56.889, 10.1063/5.0166824}. 
Nonequilibrium driven systems frequently exhibit emergent fluctuations and spatiotemporal pattern formation that cannot be inferred reliably from static observables alone~\cite{demery_driven_2019,pagare_stochastic_2019,buttinoni_active_2022,faran_nonequilibrium_2025}. 
While the states before and after a transition are often well characterized within established physical frameworks, anticipating the onset of a transition remains fundamentally more challenging~\cite{scheffer_early-warning_2009, bury_deep_2021, doi:10.1126/science.aay4895, Hock2024, https://doi.org/10.1002/2016GL068392, 10.1063/5.0158109, doi:10.1126/science.abn7950, Flores2024, 10.1063/5.0245575, MA2025119306, doi:10.1142/S0218127416502394, doi:10.1126/science.1210657, MA2019, PhysRevE.101.012206}. 
This difficulty arises because early-stage precursors are typically weak, spatially heterogeneous, and masked by intrinsic fluctuations~\cite{aryanfar_finite-pulse_2019, jang_effect_2021, wood_dendrites_2016, MA2019, https://doi.org/10.1111/j.1461-0248.2005.00877.x, scheffer_early-warning_2009, Ditlevsen2023, https://doi.org/10.1111/j.1461-0248.2008.01160.x, Radhakrishnan2025, Wang2012, doi:10.1073/pnas.2103779118}. 
Accordingly, identifying predictive indicators and early-warning signals for critical transitions has become a central objective across many complex systems~\cite{dijkstra_predictive_2021,liu_early_2024,ma_early_2026,chowdhury_adaptive_2025}.

Electrochemical dendritic deposition provides a prototypical and technologically important realization of this general problem~\cite{macfarlane_lithium-doped_1999, jacobsonContinuumLimitDendritic2025b, nielsen_sharp-interface_2015, wood_dendrites_2016, cao_lithium_2020, pathak_fluorinated_2020, chen_origin_2023, brissot_dendritic_1999}. 
The gradual evolution from compact, spatially uniform deposition to highly branched dendritic growth constitutes a nonequilibrium interfacial instability with direct consequences for device safety, performance, and lifetime~\cite{nagatani_pattern_1989, scheffer_early-warning_2009}. 
This instability originates from the competition between diffusive transport and interfacial reaction kinetics and can be characterized by a dimensionless Damk\"ohler number~\cite{cross_pattern_1993, hohenberg_dynamic_1977, PhysRevLett.56.889, brener_temkin_1995, karma_phase-field_1996, nagatani_pattern_1989},
\begin{equation}
\mathrm{Da}=\frac{kL}{D},
\end{equation}
where $k$ is an effective reaction rate, $D$ the ionic diffusivity, and $L$ a characteristic interfacial length scale. 
In the diffusion-dominated regime ($\mathrm{Da}\ll1$), incoming particles sufficiently sample the interface prior to incorporation, leading to compact growth~\cite{witten_diffusion-limited_1981, witten_diffusion-limited_1983, Meakin1998, family_vicsek_1985, barabasi_stanley_1995}. 
When reaction kinetics locally outpace diffusive relaxation ($\mathrm{Da}\gtrsim1$), interfacial perturbations are amplified, giving rise to dendritic morphologies~\cite{jacobsonContinuumLimitDendritic2025b}. 
Consequently, even for fixed reaction and transport parameters, a compact-to-dendritic transition (CDT) is generically expected as growth proceeds, although its precise onset depends sensitively on stochastic growth dynamics~\cite{somfai_scaling_2005, halsey_dla_2000, havlin_phase_1989, nauenberg_crossover_1983, ball_diffusion-controlled_1984, Ritchie16}. 
These considerations have motivated strategies such as pulsed or modulated charging protocols that attempt to delay dendritic growth by dynamically altering the effective reaction--diffusion balance~\cite{havlin_phase_1989, nauenberg_crossover_1983, ball_diffusion-controlled_1984, Ritchie16}. 

Classical frameworks such as diffusion-limited aggregation and related reaction--diffusion models successfully rationalize the morphology and scaling behavior across this transition~\cite{havlin_phase_1989, nauenberg_crossover_1983, ball_diffusion-controlled_1984}. In particular, Eden-type growth represents the reaction-limited, compact extreme, whereas diffusion-limited aggregation (DLA) corresponds to the diffusion-controlled dendritic limit~\cite{eden_1961, meakin_dla_1983}. 
However, during early growth the deposit remains globally compact, and the instability manifests only through weak, localized interfacial fluctuations~\cite{saito_muller-krumbhaar_1995, brady_fractal_1984, macfarlane_lithium-doped_1999, jacobsonContinuumLimitDendritic2025b, nielsen_sharp-interface_2015}. As a result, predicting the eventual onset of dendritic growth from early-stage configurations remains difficult, due to stochasticity and intrinsic interfacial heterogeneity~\cite{boettiger_hastings_2013, ditlevsen_tipping_2010, lenton_tipping_2008, scheffer_early-warning_2009, bury_deep_2021, masuda_anticipating_2024}. This suggests that early-warning of dendritic instability requires detecting subtle and evolving spatiotemporal correlations, rather than identifying static morphological signatures alone~\cite{aryanfar_finite-pulse_2019, jang_effect_2021, wood_dendrites_2016}. These challenges are further compounded by limited experimental access to buried interfacial dynamics, motivating the use of physics-based simulations and synthetic image data as controlled testbeds~\cite{provatas_elder_2010, brener_temkin_1995, ma_atomic_2020, wood_dendrites_2016, chen_origin_2023}.

Recent advances demonstrate that machine learning and reinforcement learning can autonomously learn effective representations and control strategies in complex physical systems, ranging from turbulent flows and active matter to self-assembly and quantum dynamics~\cite{rabault_active_flow_control_2019,duriez_ml_control_2017,klotsa_self_assembly_rl_2019,whitelam_pathway_control_2018,bukov_rl_quantum_2018,pathak_model_free_control_2018}. In interfacial growth problems, image-based learning is particularly attractive because it can implicitly encode multiscale morphological features and collective fluctuations that are difficult to capture with handcrafted descriptors~\cite{kumar_visualization-based_2022, bhattacharya_predicting_2022, PyTorch, lu_deep_2023, lecun_gradient_1998, krizhevsky_imagenet_2012}. When combined with physically grounded models, such approaches naturally lend themselves to early-warning and control in driven nonequilibrium systems, where timely intervention is required to prevent irreversible instabilities~\cite{aryanfar_finite-pulse_2019, jang_effect_2021, wood_dendrites_2016}. Related data-driven methods have also been explored for optimizing electrochemical and battery processes, including fast-charging protocols and degradation-aware control~\cite{li_battery_rl_2020,yang_fast_charging_rl_2021,wang_improving_2023}.

For example, Kumar \textit{et al.}~\cite{kumar_visualization-based_2022} developed a convolutional LSTM--based, image-driven framework to predict the spatiotemporal evolution of dendritic copper electrodeposition from real-time microscopy data, demonstrating forecasting of future growth morphologies without explicit physical modeling~\cite{nielsen_sharp-interface_2015, tikekar_stabilizing_2016, choudhury_confining_2018, vlachas_data-driven_2018, pathak_model-free_2018}. Such studies highlight the ability of deep learning to capture complex growth patterns and morphological variability~\cite{kumar_visualization-based_2022, bhattacharya_predicting_2022, wood_dendrites_2016}. However, most existing approaches focus on next-frame or future-sequence prediction and lack an explicit, physically grounded formulation of early-warning for the compact-to-dendritic transition from pre-transition configurations~\cite{jacobsonContinuumLimitDendritic2025b, Ma2026, Yu2025}.

In this work, we study compact-to-dendritic transitions (CDTs) in a particle-based electrodeposition model using image-based spatiotemporal learning combined with physics-based growth simulations~\cite{jacobsonContinuumLimitDendritic2025b}. Using trajectory-resolved transition points as ground truth, we formulate CDT forecasting as binary event prediction over a future horizon $E$ from pre-transition image sequences. By benchmarking models that disentangle or jointly learn spatial representation and temporal integration~\cite{scheffer_early-warning_2009,kuehn_critical_2011}, we show that reliable early-warning near the alarm boundary requires both effective temporal integration and end-to-end learning of joint spatial and temporal representations. Models that satisfy both (CNN--GRU and CNN--TCN) achieve consistently strong performance across wide prediction horizons, whereas approaches lacking either component degrade significantly. Analysis of the learned dynamics reveals that the CNN--GRU organizes pre-transition information into a structured latent evolution, supporting the interpretation of the hidden state as a low-dimensional surrogate for progressive morphological destabilization~\cite{vlachas_data-driven_2018,pathak_model-free_2018,masuda_predicting_2024}. Transfer across reaction-rate conditions is limited but systematic, and fine-tuning restores performance without evidence of accelerated learning. Our results establish a practical and generalizable framework for forecasting incipient morphological instabilities in driven nonequilibrium pattern-forming systems.

\begin{figure*}[htbp!]\centering
  \includegraphics[width=\textwidth]{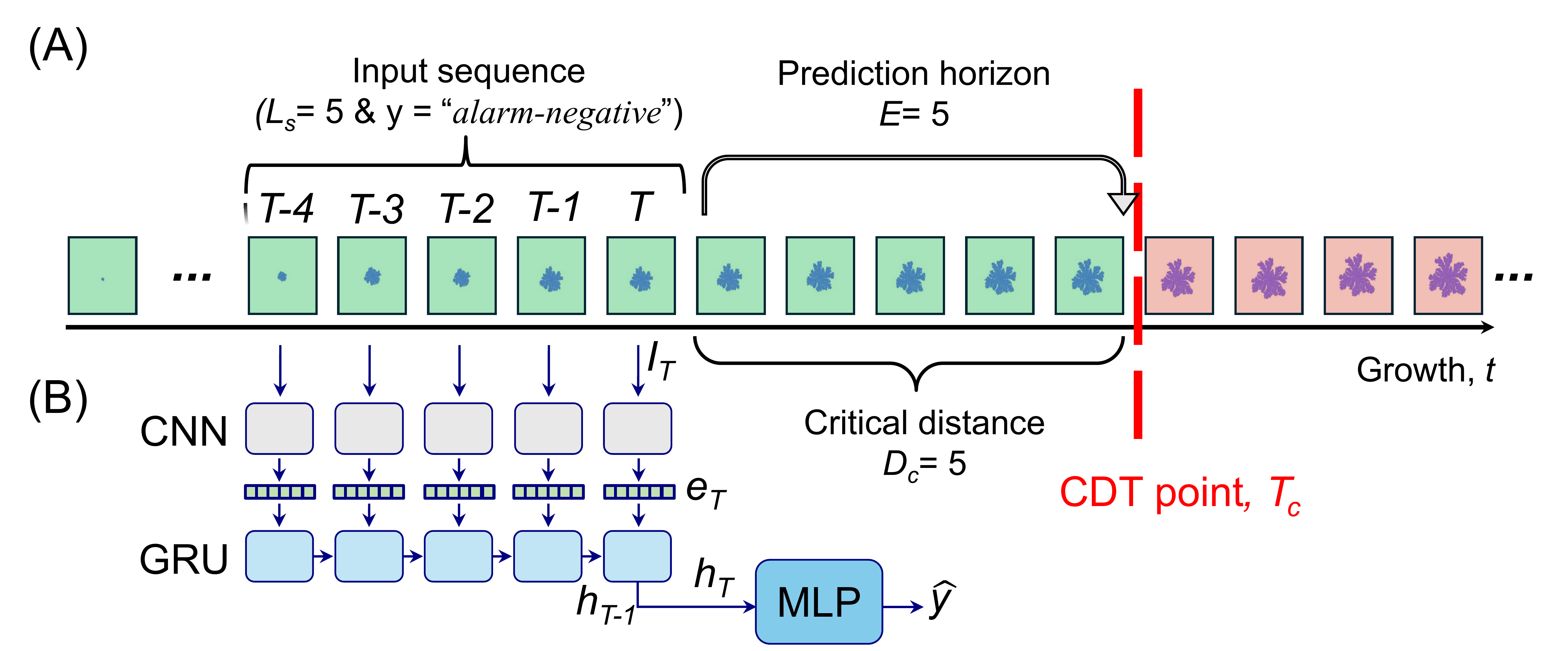}
\caption{Setup for early-warning learning.
(A) Schematic illustration of temporal sequence construction and early-warning labeling. At a reference temporal index $t=T$, the model observes an input sequence of $L_s$ past growth images, 
$\{I_{T-L_s+1}, \ldots, I_T\}$. The compact-to-dendritic transition (CDT) occurs at $T_c$, and the critical distance is defined as $D_c = T_c - T$. The label $y$ is assigned as alarm-positive if $T_c \in [T,\, T+E)$, i.e., if the transition occurs within a future prediction horizon of length $E$, and as alarm-negative otherwise.
(B) Architecture of the CNN--GRU model for end-to-end spatiotemporal learning. CNN extracts spatial feature representations $e_t$ from each image, GRU models their temporal evolution into a hidden state $h_T$, and an MLP outputs the predicted label $\hat{y}$ (alarm-positive or alarm-negative).}
\label{fig:early-alarm-illustration}
\end{figure*}

\section{Particle-based model and compact-to-dendritic transition}

We model electrodeposition in two dimensions using a particle-based aggregation model that captures the competition between diffusion and surface reaction, a minimal description known to reproduce both compact and dendritic growth morphologies~\cite{jacobsonContinuumLimitDendritic2025b}. Growth proceeds through repeated cycles of Brownian diffusion of reactive particles in the electrolyte domain followed by probabilistic sticking upon contact with the growing aggregate, representing electrochemical reduction at the interface. The reaction probability is controlled by a rate constant $k$, which interpolates between reaction-limited (compact) and diffusion-limited (dendritic) growth. All simulations are performed using a publicly available implementation with parameters identical to those in Ref.~\cite{DanGithub}. Further algorithmic details of the reaction-diffusion model are provided in Appendix~\ref{sec:appendix:particle}.

\subsection{Compact-to-dendritic transition (CDT)}\label{sec:cdt}

The fractal dimension $d_f$ provides a quantitative measure of aggregate morphology through the scaling relation
\begin{equation}\label{eq:fractal}
  N(R) \sim R^{d_f},
\end{equation}
where $N(R)$ denotes the number of deposited particles contained within a distance $R$ from the aggregate center. Physically, values of $d_f \approx 2$ correspond to compact, space-filling growth, whereas lower values ($d_f \approx 1.7$) are characteristic of branched or dendritic morphologies~\cite{havlin_phase_1989, nauenberg_crossover_1983, ball_diffusion-controlled_1984}. The evolution of $d_f$ therefore serves as a sensitive indicator of structural changes during aggregation. We compute $d_f$ numerically following Ref.~\onlinecite{jacobsonContinuumLimitDendritic2025b}:
\begin{equation}\label{eq:df}
  d_f = \frac{d \log_{10} N}{d \log_{10} (k R_g / D)},
\end{equation}
where $R_g$ is the radius of gyration of the aggregate. This differential definition captures local deviations from global scaling and is therefore well suited for detecting morphological transitions during nonequilibrium growth for each growth trajectory.

The onset of dendritic growth for each trajectory is identified by a critical radius $R_c$,
\begin{equation}\label{eq:rc_def}
  R_c = \arg\max_{R_g}\left(\frac{d \log_{10} N}{d \log_{10} (k R_g / D)}\right),
\end{equation}
which marks the CDT and follows the scaling relation $R_c \sim D/k$. Across this CDT transition, the effective fractal dimension $d_f$ decreases gradually, reflecting a morphological crossover from compact to dendritic growth.
For image-based analysis, instead of the ensemble-averaged value of $R_c$, we compute a sequence-specific $R_c$ for each growth trajectory. This trajectory-resolved definition serves as a ground-truth reference for subsequent compact-dendritic classification. Some figures show the ensemble-averaged $R_c$ as a visual reference unless noted. We note that accurate determination of the ensemble-averaged $R_c$ requires extensive ensemble averaging over many independent growth trajectories, which is not the primary objective of this work. 

In addition to the fractal dimension $d_f$, other physical quantities provide complementary geometric measures of morphological evolution across the CDT~\cite{jacobsonContinuumLimitDendritic2025b} (see Appendix~\ref{sec:appendix:images_phys} for calculation details). For instance, the interfacial length $L_{\mathrm{int}}$ characterizes the degree of interfacial roughness and branching of the deposited aggregate and is defined as the perimeter of the aggregate boundary in two dimensions. In the compact growth regime prior to the CDT, the interface remains relatively smooth and $L_{\mathrm{int}}$ scales linearly with aggregate size, $L_{\mathrm{int}} \sim R_g$~\cite{aryanfar_finite-pulse_2019, jang_effect_2021, wood_dendrites_2016}. Beyond the transition, the emergence of branching and interfacial instabilities leads to a nontrivial scaling,
\begin{equation}
  L_{\mathrm{int}} \sim R_g^{d_i},
\end{equation}
where the interfacial exponent $d_i$ is empirically found to lie in the range $1.66 \lesssim d_i \lesssim 1.7$~\cite{Meakin1998}, reflecting the increased geometric complexity of the dendritic interface. To obtain a dimensionless quantity with behavior analogous to $d_f$, we define the normalized interfacial length as
\begin{equation}\label{eq:inter_norm}
  \tilde{L}_{\mathrm{int}} = \frac{L_{\mathrm{int}} R_g}{A},
\end{equation}
where $A$ is the cluster area. This quantity increases in the dendritic post-CDT regime.

\begin{figure*}[htbp!]\centering
\includegraphics[width=5.5in]{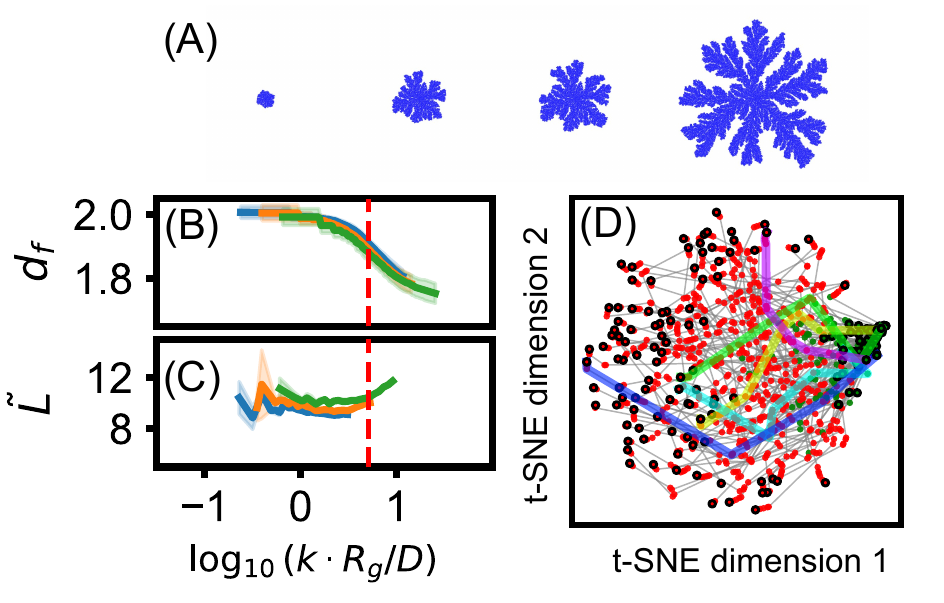}
\caption{Morphological characterization across the compact-to-dendritic transition (CDT).
(A) Representative electrodeposition morphologies during growth at $\log_{10} k = -1.69$.
(B) Fractal dimension $d_f$ (Eq.~\ref{eq:df}) as a function of growth progression.
(C) Dimensionless interfacial length  $\tilde{L}_{\mathrm{int}}$ (Eq.~\ref{eq:inter_norm}) as a function of growth progression.
Panels (B) and (C) present results for three reaction rates, $\log_{10} k = -1.69$ (green), $-1.90$ (orange), and $-2.12$ (blue). As a visual reference, red dashed lines indicate the CDT at $R_g = R_c$. Here, the ensemble-averaged value of $R_c$ is adopted, reported in Ref.~\onlinecite{jacobsonContinuumLimitDendritic2025b}.
(D) Two-dimensional t-SNE embedding of 100 growth trajectories at $\log_{10} k = -1.69$ constructed from the $e_{\mathrm{phys}}$ representations. Markers with black edges denote the starting and end points of each trajectory, and marker color indicates the morphological state: green for compact growth and red for dendritic growth. Grey lines connect each trajectory. Thick colored lines show five representative growth trajectories in the latent space.} \label{fig:cdt_morph}
\end{figure*}

\section{Early-warning task and models}

Particle configurations from deposition simulations are mapped onto binary images $I_t \in \mathbb{R}^{H \times W}$ at regular growth intervals corresponding to $\Delta N = 10^3$ newly deposited particles, thereby defining an updated temporal index $t$. This procedure yields $10^3$ temporal image frames along each growth trajectory. We emphasize that the index $t$ represents the growth progression in the strict sense, not directly the physical time of the evolving system, which is not tracked in the deposition simulations employing algorithmic techniques for efficient sampling~\cite{jacobsonContinuumLimitDendritic2025b}. Nevertheless, the early-warning framework developed in this work can be straightforwardly applied to any genuinely time-evolving system. Unless otherwise stated, the results presented here focus on the case $H \times W = 32 \times 32$, for which the interfacial morphologies remain clearly identifiable (see Fig.~\ref{si:fig:grid}). Further details of the image processing pipeline, including image construction, normalization, and scaling procedures, are provided in Appendix~\ref{sec:appendix:images_pre}.

The early-warning task is formulated as binary event-prediction on the future morphology of temporal sequences. At temporal index $t=T$, the model observes an input window $\{I_{T-L_s+1},\ldots,I_T\}$ and predicts whether the CDT will occur within a future horizon of length $E$. An input sequence, whose images all belong to the pre-CDT regime, is labeled \textit{alarm-positive} if $T_c\in[T,T+E)$ and \textit{alarm-negative} otherwise, where $T_c$ is the critical time when $R_g=R_c$. Equivalently, defining $D_c=T_c-T\ge0$, sequences are alarm-negative for $D_c\ge E$ and alarm-positive otherwise. No input sequence used for prediction contains images from the post-CDT regime.  The triplet $(L_s,E,D_c)$ characterizes historical context, prediction horizon, and earliness. All models are trained and evaluated using identical $(L_s,E)$ and data splits. We stress that, because the label assignment explicitly depends on the prediction horizon $E$, a separate model must be trained for each value of $E$. Otherwise, an identical input sequence of length $L_s$ could be labeled as either alarm-positive or alarm-negative, leading to an ill-posed learning problem. 

To prevent temporal leakage, data splitting is performed at the level of independent simulation runs rather than individual sequences. The full dataset is first partitioned into a training set ($80\%$) and a held-out test set ($20\%$). The training set is then further divided into five folds with an internal $4{:}1$ ratio for training and validation, enabling five-fold cross-validation, corresponding to approximately $64\%$ training, $16\%$ validation, and $20\%$ testing data overall. Stratification is applied to preserve class balance. 
In total, we use 200 independent growth trajectories, each consisting of 1000 images.  Final performance metrics are obtained by evaluating each fold-trained model and reporting the average across folds. For all models tested, learning curves for both training and validation decreases well within 30 epochs.

\subsection{CNN--GRU architecture} \label{sec:cnn_gru_model}

We employ a hybrid CNN--GRU architecture that learns spatiotemporal representations directly from raw image sequences in an end-to-end manner~\cite{wang_artificial_2020, donahue2016longtermrecurrentconvolutionalnetworks}. The model explicitly separates spatial pattern extraction, handled by a convolutional neural network (CNN)~\cite{lecun_gradient_1998}, from temporal dependency modeling, handled by gated recurrent units (GRUs)~\cite{cho2014learningphraserepresentationsusing}. This separation reflects the multiscale nature of dendritic growth dynamics, in which localized morphological patterns evolve over time and collectively give rise to macroscopic instability. The overall model $f_\theta$ can be written as
\begin{equation}
  f^{\mathrm{CNN-GRU}}_\theta(I_t)
  = \mathrm{MLP}\!\left(\mathrm{GRU}\!\left(\mathrm{CNN}(I_t)\right)\right),
\end{equation}
where $\theta$ denotes the set of learnable parameters, and the multilayer perceptron (MLP) is used for final binary event-prediction. 

Each frame $I_t$ is processed independently by a CNN encoder,b
\begin{equation}
  F_t = \mathrm{CNN}(I_t) \in \mathbb{R}^{S_F \times (H_F \times W_F)},
\end{equation}
where the encoder consists of three convolutional blocks followed by adaptive average pooling to a fixed spatial resolution. The resulting feature map $F_t$ contains $S_F$ feature channels resolved over a $H_F \times W_F$ grid of coarse spatial bins. Unless noted otherwise, the main results in this work use $H_F \times W_F = 1 \times 1$ and $S_F = 16$, corresponding to global spatial pooling that emphasizes feature-type information over explicit spatial localization. Flattening yields a frame-level embedding $e_t = \mathrm{vec}(F_t)$.

The sequence of embeddings $\{e_t\}_{t=T-L_s+1}^{T}$ is then fed into a gated recurrent unit,
\begin{equation}
  h_t = \mathrm{GRU}(h_{t-1}, e_t) \in \mathbb{R}^{H_d},
\end{equation}
where $H_d$ denotes the hidden-state dimension. The GRU produces a sequence of hidden states $H = \{h_t\}_{t=T-L_s+1}^{T}$, from which the final hidden state $h_T$ is used as a compact summary of the input sequence. This final hidden state is mapped to a scalar logit,
\begin{equation}
  \hat{y} = \mathrm{MLP}(h_T),
\end{equation}
which represents the unnormalized confidence for the alarm-positive class.

The CNN--GRU model is trained end-to-end using a weighted binary cross-entropy loss~\cite{He2009} with logits, implemented as \texttt{BCEWithLogitsLoss} in \texttt{PyTorch} (version 2.4.1)~\cite{PyTorch}. Optimization is performed with the \texttt{AdamW} optimizer~\cite{loshchilov2019decoupledweightdecayregularization}. Regularization includes batch normalization~\cite{ioffe2015batchnormalizationacceleratingdeep}, dropout~\cite{JMLRdropout}, gradient clipping~\cite{pmlr-v28-pascanu13}, and early stopping~\cite{earlystopping}, together with a \texttt{ReduceLROnPlateau} scheduler for adaptive learning-rate control. PyTorch implementations of all models investigated in this work are publicly available at Ref.~\onlinecite{HJGitHub}.

\subsection{Alternative spatial and temporal encoding models}

Regarding spatial encoding, in addition to the CNN-based representation, we employ a complementary encoding motivated by the physical characterization of deposition growth (Eq.~\ref{eq:fractal}). For each time frame $t$, the physics-informed encoder, denoted $\mathrm{Rad}^{\mathrm{phys}}$, computes the radial density gradient from the center of the image $I_t$. This procedure yields a physics-based feature vector $e^{\mathrm{phys}}_t$ of dimension $F$~\cite{jacobsonContinuumLimitDendritic2025b, wang_artificial_2020},
\begin{align}\label{eq:Fphys}
  e^{\mathrm{phys}}_t &= \mathrm{Rad}^{\mathrm{phys}}(I_t), \\
  &= \frac{d}{dR}\bigg[\frac{N(R)}{\pi R^2}\bigg] \in \mathbb{R}^{F}.
\end{align}
Similar to the CNN encoder, this representation coarsens the underlying spatial information; however, it does so through explicitly defined, physically interpretable quantities rather than learned convolutional filters. We note that $e^{\mathrm{phys}}_t$ may also be constructed from alternative physically motivated descriptors beyond those considered in this work.

To assess the role of temporal integration in the CNN--GRU framework, we consider several alternative temporal models. As non-temporal baselines, we employ a linear regressor ($\mathrm{LR}$)~\cite{inbook} and a MLP~\cite{Rumelhart1986LearningRB, Goodfellow-et-al-2016}, both of which operate on the same frame-level input features but lack explicit mechanisms for integrating information across time. The linear regressor provides a purely linear reference, whereas the MLP allows for nonlinear feature transformations without modeling temporal dependencies.

In addition, we consider a temporal convolutional network (TCN)~\cite{bai_empirical_2018, tcn_energy_forecasting_2020, bednarski_tcn_clinical_2022} as an alternative deep-learning model for temporal integration. Unlike recurrent architectures that accumulate information through hidden-state memory, the TCN aggregates temporal context via convolutional filtering along the time axis~\cite{aryanfar_finite-pulse_2019, jang_effect_2021, wood_dendrites_2016}. This comparison enables us to distinguish whether the performance gains of the CNN--GRU primarily arise from recurrent memory or from temporal integration more generally, independent of the specific architectural realization. All models tested in this work are summarized in Table~\ref{tab:model_summary}.

\begin{table}[tbp]
\small
\begin{tabular*}{\linewidth}{lccc}
\hline
\textbf{Model} & \textbf{Spatial} & \textbf{Temporal} & \textbf{Notes} \\
\hline
CNN--GRU & CNN & GRU & End-to-end spatiotemporal \\
CNN--TCN & CNN & TCN & End-to-end spatiotemporal\\
CNN$^{cl}$--GRU & CNN$^{cl}$ & GRU & Fixed spatial \\
CNN$^{cl}$--LR & CNN$^{cl}$ & LR & Linear static baseline \\
CNN$^{cl}$--MLP & CNN$^{cl}$ & MLP & Nonlinear static baseline \\
Rad$^{phys}$--GRU & Rad$^{phys}$ & GRU & Physics-informed spatial \\
Rad$^{phys}$--MLP & Rad$^{phys}$ & MLP & Physics-informed spatial \\
CNN--LR & CNN & LR & End-to-end \\
CNN--MLP & CNN& MLP & End-to-end \\
\hline
\end{tabular*}
\caption{Summary of model architectures and their spatial and temporal components.}
\label{tab:model_summary}
\end{table}

\section{Results and Discussion}
In this section, we present results for early-warning prediction of CDTs. We first characterize morphological signatures of the transition using physics-based descriptors, which serve as ground truth for the prediction task. We then quantitatively compare the performance of early-warning models, highlighting the necessity of end-to-end spatiotemporal learning for reliable prediction near the CDT. Next, we analyze the learned latent  hidden-state dynamics of the CNN--GRU to assess their interpretation as low-dimensional surrogates for morphological evolution. Finally, we examine the transferability of the learned representations across different reaction-rate conditions, which turns out to be limited but systematic.

\begin{figure*}[htbp!]\centering
  \includegraphics[width=\textwidth]{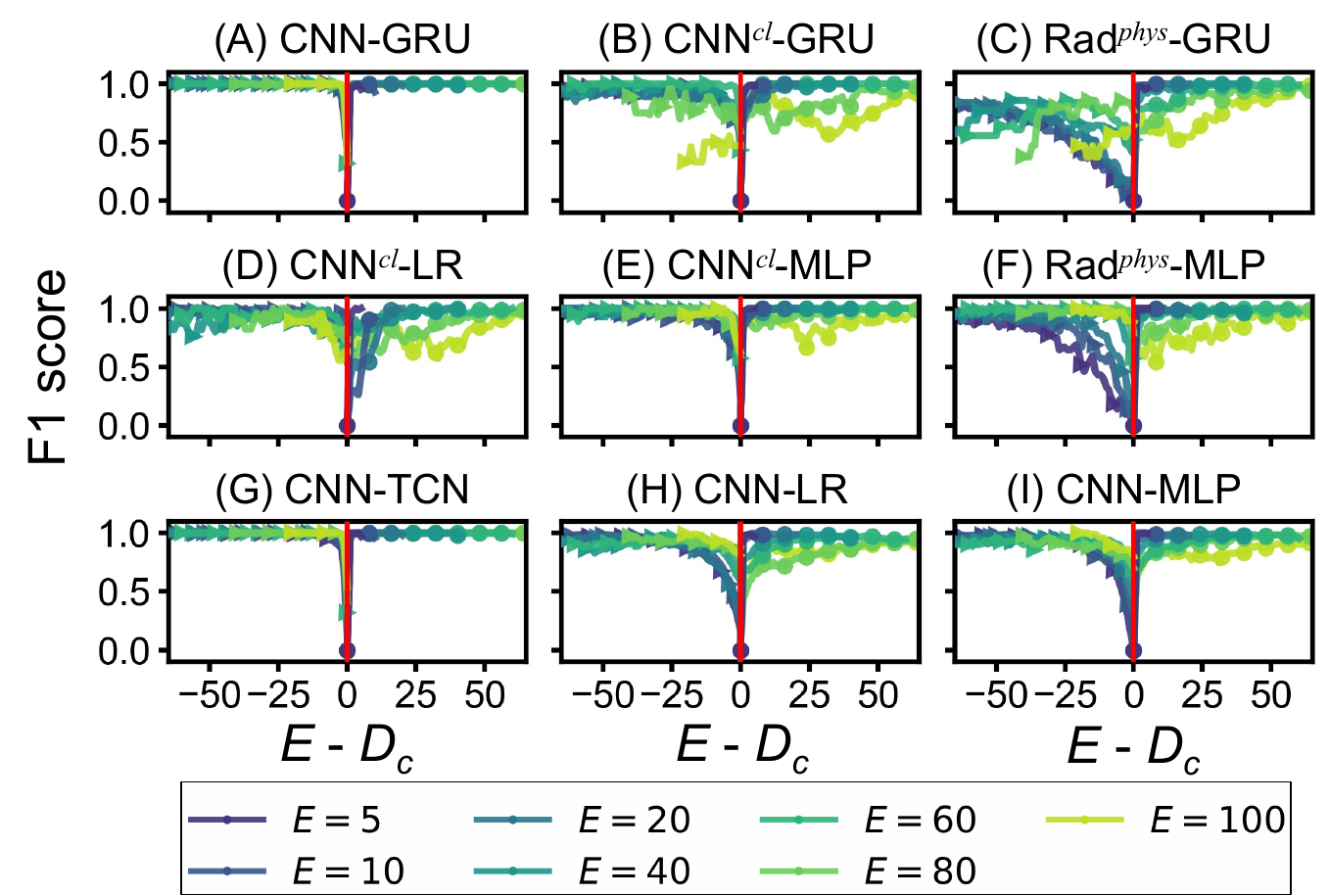}
\caption{
F1-score performance of early-warning models as a function of $E - D_c$ at $L_s = 5$. Circle and triangle markers denote the F1-positive and F1-negative scores, respectively (Eq.~\ref{eq:f1_both}). Here, $D_c = T_c - T$ is the temporal distance between the prediction time and the trajectory-specific transition point, and $E$ is the prediction horizon. $E \leq D_c$ corresponds to the alarm-negative (compact) regime, whereas $E > D_c$ corresponds to the alarm-positive regime. $D_c$ is determined using the sequence-specific $R_c$.
Panels A, G, H, and I correspond to end-to-end learning models.
Red vertical lines denote the boundaries between the two early-warning prediction regimes. The grid resolution ($H \times W = 32 \times 32$), CNN feature dimension ($S_F = 16$), and global spatial pooling size ($H_F \times W_F = 1 \times 1$) are fixed across all models.
} \label{fig:main_results}
\end{figure*}

\subsection{Identification of morphological signatures across the CDT}

We first examine how compact-to-dendritic transitions (CDTs) manifest in physics-based morphological descriptors and physics-derived feature representations (Fig.~\ref{fig:cdt_morph}). This analysis establishes physically grounded signatures of the transition that serve as a reference for subsequent early-warning prediction and model comparison, while also highlighting the pronounced heterogeneity across individual growth trajectories.

A growing deposit undergoes a morphological crossover from compact to dendritic growth across the CDT (Fig.~\ref{fig:cdt_morph}A). To quantify this crossover, we consider the fractal dimension $d_f$, which provides a robust and sensitive measure of morphological evolution across the transition (Fig.~\ref{fig:cdt_morph}B). Consistent with previous studies~\cite{jacobsonContinuumLimitDendritic2025b}, an appropriate renormalization of the growth variables yields a unified description of the CDT across different reaction rates. For all values of $k$ considered ($\log_{10} k \in \{-2.12, -1.90, -1.69\}$), growth initially proceeds in a compact manner and subsequently transitions to a dendritic morphology beyond a critical radius $R_c$~\cite{jacobsonContinuumLimitDendritic2025b}. The location of the transition depends systematically on $k$, with higher reaction rates leading to earlier transitions and dendritic growth emerging at smaller aggregate sizes. We again note that the value of diffusion coefficient $D$ is held fixed throughout this work.

As shown in Fig.~\ref{fig:cdt_morph}C, the dimensionless interfacial length $\tilde{L}_{\mathrm{int}}$ also exhibits a gradual crossover across the CDT. It responds sensitively to the regime near and beyond $R_c$ through an increase in magnitude, reflecting the emergence of ramified, dendritic interfaces relative to compact growth. Other physics-based descriptors, including circularity, aspect ratio, and convex hull ratio, display similar trends (Fig.~\ref{si:fig:phys} in the Appendix); however, their variations across $R_c$ are comparatively smooth and strongly correlated, limiting their ability to unambiguously localize the transition at the level of individual trajectories.

Despite the well-defined $R_c$ observed at the ensemble-averaged level, individual growth trajectories exhibit substantial heterogeneity due to the inherently stochastic nature of the deposition process. This heterogeneity is illustrated in Fig.~\ref{fig:cdt_morph}D, which shows a two-dimensional t-SNE embedding of the physics-derived feature vectors $e_{\mathrm{phys}}$ (Eq.~\ref{eq:Fphys}) constructed from all trajectories. No learning is involved in this representation, as $e_{\mathrm{phys}}$ consists solely of hand-crafted, physics-motivated descriptors. While trajectories remain relatively clustered in the compact regime, a pronounced spread emerges in the dendritic regime, reflecting the diversity of post-CDT morphologies. Such variability complicates the assignment of morphological states based solely on an ensemble-averaged $R_c$. Accordingly, we employ trajectory-resolved transition points $R_c$ (Eq.~\ref{eq:rc_def}) as ground truth labels for the supervised early-warning models considered below.

\begin{figure}[htbp!]\centering
  \includegraphics[width=3.2in]{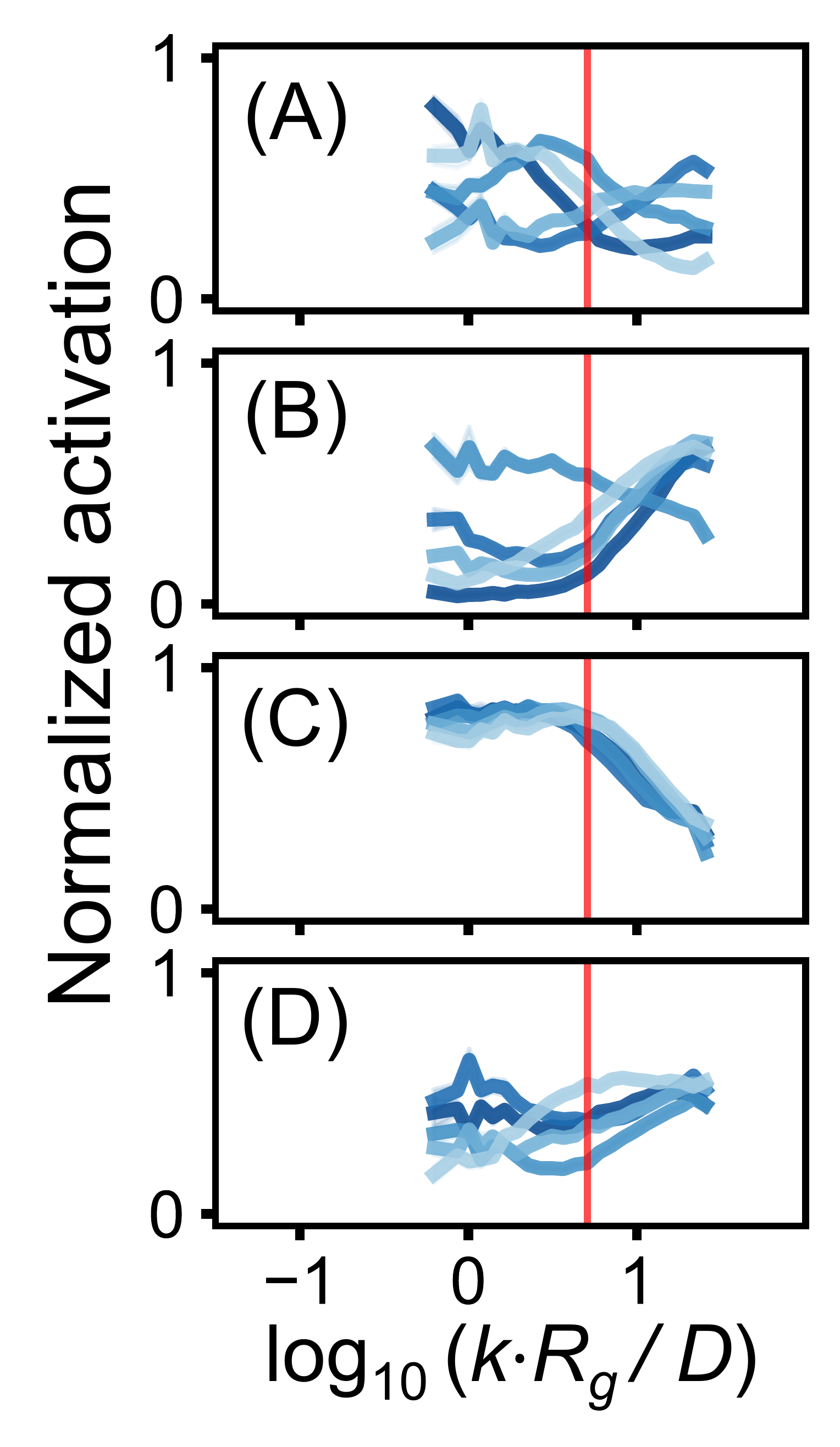}
  \caption{Normalized activation of the five most important spatial features $e_t$ for $E=5$ during growth at $\log_{10} k=-1.69$: (A) CNN--GRU, (B) CNN$^{cl}$, (C) Rad$^{phys}$, and (D) CNN--TCN. As a visual reference, red vertical lines indicate the CDT at $R_g = R_c$, adopting the ensemble-averaged value of $R_c$ reported in Ref.~\onlinecite{jacobsonContinuumLimitDendritic2025b}. Bluer colors indicate features with higher importance score.}
  \label{fig:embedding}
\end{figure}

\subsection{The CNN--GRU enables early-warning prediction of the CDT}

The early-warning task evaluates how reliably each method can anticipate the onset of dendritic growth given an observation window of length $L_s$ and a future prediction horizon $E$, which together define the temporal distance $D_c$ between the prediction time and the sequence-specific transition time $T_c$ (Fig.~\ref{fig:early-alarm-illustration}). Performance is quantified using two types of F1 scores~\cite{Blair1979InformationR2}: $\mathrm{F1_{positive}}$ and $\mathrm{F1_{negative}}$, defined as
\begin{equation}\label{eq:f1_both}
\begin{aligned}
\mathrm{F1_{positive}} &= \frac{\mathrm{TP}}{\mathrm{TP} + \tfrac{1}{2}(\mathrm{FP} + \mathrm{FN})},\\[5pt] 
\mathrm{F1_{negative}} &= \frac{\mathrm{TN}}{\mathrm{TN} + \tfrac{1}{2}(\mathrm{FN} + \mathrm{FP})}.
\end{aligned}
\end{equation}
Here, $\mathrm{TP}$, $\mathrm{FP}$, $\mathrm{TN}$, and $\mathrm{FN}$ denote the numbers of true positives, false positives, true negatives, and false negatives, respectively. For each value of $E$, the F1 scores are computed over the ensemble of test sequences.

Figure~\ref{fig:main_results} summarizes both F1 scores (Eq.~\ref{eq:f1_both}) for binary forecasting of growth trajectories at $\log_{10} k=-1.69$ with fixed $L_s=5$, plotted as a function of $E-D_c$. In this representation, $D_c$ specifies the location of the true transition relative to the prediction time, while $E-D_c$ determines the expected label: $E-D_c<0$ corresponds to the alarm-negative regime, $E-D_c>0$ to the alarm-positive regime, and $E-D_c=0$ marks the alarm-boundary. By construction, $\mathrm{F1_{positive}}$ primarily reflects performance in the alarm-positive regime, whereas $\mathrm{F1_{negative}}$ emphasizes performance in the alarm-negative regime. Although the sequence label (alarm) depends only on the sign of $E-D_c$, the conditional distribution of pre-transition morphologies at fixed $E-D_c$  changes with $E$ because of different temporal proximity-to-CDT.

All models exhibit better performance in temporal regions far from $E \approx D_c$ than in the vicinity of $E \approx D_c$ in both regimes, consistent with the fact that the early-warning task is most challenging near the CDT, where morphological precursors are weak and buried by intrinsic fluctuations. For the same value of $E-D_c$, the F1 performance further depends on $E$: in the alarm-negative regime relative to $E$, prediction is relatively easy for small $E$, whereas in the alarm-positive regime relative to $E$, prediction becomes easier for large $E$.

Beyond these general trends, the F1 performance exhibits clear model-specific differences, highlighting the importance of simultaneous spatiotemporal learning together with effective temporal information integration. The CNN--GRU model (Fig.~\ref{fig:main_results}A) consistently outperforms all other approaches, with the CNN--TCN (Fig.~\ref{fig:main_results}G) achieving comparable F1 scores. Notably, in the most challenging region near the alarm-boundary, only the CNN--GRU and CNN--TCN maintain high F1 scores across a broad range of prediction horizons. This indicates that effective temporal integration is not limited to a specific recurrent-memory mechanism.

Models that incorporate temporal modeling but rely on fixed or separately learned spatial representations (CNN$^{cl}$--GRU and Rad$^{phys}$--GRU; see Figs.~\ref{fig:main_results}B--C) underperform their LR and MLP counterparts and remain systematically inferior to the CNN--GRU. For the same temporal modeling, CNN$^{cl}$ outperforms Rad$^{phys}$ in the vicinity of $E=D_c$ in both regimes. Rad$^{phys}$ tends to suffer in detecting challenging cases near $T_c$, misclassifying a late-stage compact morphology as alarm-positive and an early-stage dendritic morphology as alarm-negative. Consequently, Rad$^{phys}$ exhibits high false alarms for both positive and negative when $E\approx D_c$. Regarding temporal learning alone, MLP outperforms LR, as expected, primarily by reducing false-negative alarms when $E\gtrsim D_c$. These results demonstrate that temporal modeling alone is insufficient without jointly optimized spatial representations.

Models that employ simultaneous spatiotemporal learning without effective temporal integration (Figs.~\ref{fig:main_results}H--I) also show degraded performance, particularly in the vicinity of $E=D_c$ in both regimes (see Fig.~\ref{si:fig:f1_zoom} in the Appendix for a magnified view). Therefore, reliable early-warning with a wide prediction horizon near the CDT requires the co-implementation of two essential components: (i) temporal integration and (ii) simultaneous end-to-end learning of spatial representations. The likelihood results before thresholding further support this conclusion, showing the overwhelming performance of both CNN--GRU and CNN--TCN models (see Fig.~\ref{si:fig:likelihood} in the Appendix).

We further examine the effect of spatial resolution in the CNN--GRU model by varying the CNN feature size ($S_F$) and the degree of spatial pooling ($H_F\times W_F$) (Figs.~\ref{si:fig:f1_var} and~\ref{si:fig:likelihood_var} in the Appendix). While all configurations achieve strong F1-score performance, increasing spatial detail through larger $S_F$ or $H_F\times W_F$ leads to system-dependent changes in the likelihood results. Notably, these modifications do not systematically enhance early-warning prediction and, in some cases, degrade performance as reflected in the likelihood analysis. For example, for $S_F=16$ (as in Fig.~\ref{fig:main_results}), the likelihood remains similar or even deteriorates as $H_F\times W_F$ increases. This nonmonotonic behavior indicates a trade-off between preserving fine spatial detail and maintaining a compact, noise-robust representation.

\subsection{Spatiotemporal learning with CNN--GRU provides early-warning signals of the CDT}

The importance of simultaneous spatiotemporal learning is further underscored by the temporal evolution of the learned spatial features $e_t$. For interpretability, we additionally compute activations and latent trajectories over the full growth trajectory (including post-CDT frames). Figures~\ref{fig:embedding} display the normalized activation $a_t \in [0,1]$ of the five most important spatial features $e_t$ for $E=5$, obtained via min--max normalization (see Appendix ~\ref{sec:appendix:activ_norm} for details). All models exhibit changes in feature activations across the CDT, but their characteristic behaviors and predictive utility differ. We note that the activation patterns exhibit only weak dependence on the prediction horizon $E$ (see Fig.~\ref{si:fig:activation_grid} in the Appendix).

The physics-derived features $e_{\mathrm{phys}}$ exhibit activations that remain nearly constant in the pre-CDT regime and then drop rapidly after $R_c$ (Fig.~\ref{fig:embedding}C). This behavior closely resembles the evolution of the fractal dimension $d_f$, which likewise shows little sensitivity prior to $R_c$. Similarly, the feature activations of CNN$^{cl}$ closely mirror the behavior of $\tilde{L}_{\mathrm{int}}$ (Fig.~\ref{fig:embedding}B): they remain near zero in the pre-CDT regime and increase sharply only after $R_c$. This reflects that CNN$^{cl}$ is trained for static compact--dendritic classification and therefore primarily responds to more developed post-transition morphologies. Such ReLU-like responses are insufficiently structured to provide reliable early-warning capability in the regime of weak precursors prior to the CDT.

By contrast, the CNN--GRU exhibits qualitatively distinct activation patterns (Fig.~\ref{fig:embedding}A). A subset of features evolves gradually within the alarm-negative regime, indicating sensitivity to progressive morphological destabilization, while other features display pronounced peaks prior to $R_c$, signaling the impending transition. We interpret these pre-transition peaks as being suggestive of the emergence of localized interfacial instabilities that are insufficient to produce a globally dendritic morphology but nevertheless indicate a loss of compact-growth stability. Such precursors are weak, spatially localized, and strongly history-dependent, rendering them difficult to identify from single-frame descriptors. Their accumulation through temporal integration provides a natural explanation for the superior early-warning capability of the CNN--GRU \cite{scheffer_early-warning_2009}. Moreover, the relatively uniform importance across $e_t$ features supports that early morphological precursors of the CDT are not encoded in a single dominant spatial motif, but are instead distributed over multiple subtle and heterogeneous local patterns, consistent with the need for a distributed spatiotemporal representation.

Saliency analysis provides complementary qualitative insight into the physical content of the learned spatial features by highlighting input regions that most strongly influence the model predictions~\cite{simonyan2014deep,zeiler2014visualizing,selvaraju2017grad}. Near $R_c$, the most salient pixels are predominantly concentrated along the deposit interface, with additional contributions in the surrounding region, indicating sensitivity to both interfacial morphology and aggregate extent (see Fig.~\ref{si:fig:gbp} in the Appendix). These results further support that the spatial features encode physically meaningful aspects of the evolving growth morphology. 

While individual feature activations differ markedly between the models, the global geometric organization of trajectories in the embedding space is similar: the two-dimensional t-SNE embeddings of growth trajectories constructed from $e_t$ closely resemble the structure observed for the physics-derived representation $e_{\mathrm{phys}}$ (see Fig.~\ref{si:fig:tsne_all}~A in the Appendix). In all latent spaces, growth trajectories form a compact-regime core and subsequently radiate into a heterogeneous dendritic manifold, reflecting the stochastic and history-dependent nature of dendritic growth. We note that, for the CNN--GRU, the global spatial pooling resolution ($H_F\times W_F$) at a given feature dimension ($S_F$) can alter the latent-space manifolds, as it controls the amount of spatial information retained in $e_t$. In general, finer spatial decomposition tends to yield improved separation between compact and dendritic manifolds in the t-SNE space (see Fig.~\ref{si:fig:tsne_all} in the Appendix).

\begin{figure}[htbp!]\centering
  \includegraphics[width=3.5in]{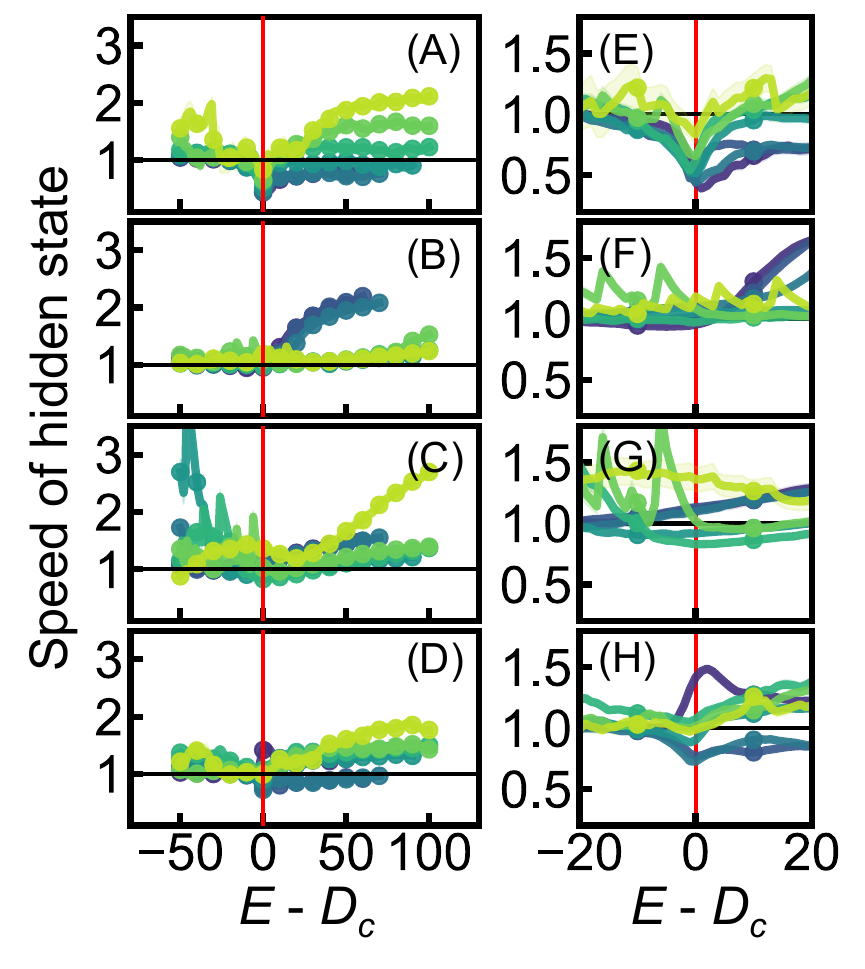}
\caption{Latent-space speed during growth at $\log_{10} k = -1.69$.
(A--C) Hidden-state velocity $v_t^h = \lVert h_t - h_{t-1} \rVert$ for the GRU-based models:
(A) CNN--GRU,
(B) CNN$^{cl}$--GRU, and
(C) Rad$^{phys}$--GRU.
(D) Latent speed $v_t^C = \lVert c_t - c_{t-1} \rVert$ for the CNN--TCN model, computed from the feature vector $c_t$ of the final hidden layer.
(E--H) Magnified views of the regions near the transition point $T_c$ for panels (A--D), respectively.
Different colors indicate different values of $E$, consistent with the color coding in Fig.~\ref{fig:main_results}. Red vertical lines denote the boundaries between the two early-warning prediction regimes, and black horizontal lines serve as guides to the eye.
} \label{fig:hidden}
\end{figure}

\subsection{Hidden state as a surrogate for morphological evolution}

To test whether it serves as a surrogate for the evolving morphological state, we examine the latent-space dynamics of the GRU hidden state $h_t$ , in line with prior work showing that learned latent representations can compactly encode and forecast complex physical dynamics~\cite{vlachas_data-driven_2018,pathak_model-free_2018,masuda_predicting_2024}. Figure~\ref{fig:hidden} shows the magnitude of the hidden-state velocity, $v_t^h = \lVert h_t - h_{t-1} \rVert$, during growth for three GRU-based models: CNN--GRU, CNN$^{cl}$--GRU, and Rad$^{phys}$--GRU. We also include the CNN--TCN, which has no recurrent hidden state. In this case, we compute an analogous latent speed, $v_t^c = \lVert c_t - c_{t-1} \rVert$, using the feature vector $c_t$ from the final hidden layer of the TCN, which plays a role analogous to that of $h_t$ in the GRU. Results are displayed as a function of $E-D_c$, consistent with the early-warning performance shown in Fig.~\ref{fig:main_results}.

Similar to the activation behavior of $e_t$, the latent-space speed exhibits markedly model-specific trends. For the CNN--GRU, the hidden-state dynamics shows pronounced changes near the alarm-boundary, spanning both alarm-positive and alarm-negative regions for all investigated values of $E$ (Fig.~\ref{fig:hidden}A). This behavior closely mirrors the trends observed in the activations of the learned spatial features, which also capture signatures in both regimes. It is notable that a reduction of $v_t^h$ upon approaching the transition from either regime is suggestive of a phenomenology qualitatively reminiscent of slowing-down observed in some complex systems~\cite{nagatani_pattern_1989, scheffer_early-warning_2009} (Fig.~\ref{fig:hidden}E). 

Moreover, in the alarm-negative regime, $v_t^h$ collapses onto a single curve for most values of $E$, whereas it diverges in the alarm-positive regime. This behavior is consistent with the structure observed in the t-SNE embeddings. The resulting $E$-dependent divergence in the post-CDT regime implies that capturing local morphological features requires increasingly fine spatial resolution, reflecting the strong heterogeneity of dendritic morphologies across growth trajectories. Consistent behavior is observed upon varying $H_F\times W_F$: finer spatial bins lead to enhanced collapse in the post-CDT regime, indicating improved capture of local morphological evolution (See Fig.~\ref{si:fig:hidden_var} in the Appendix). 

In contrast, the CNN$^{cl}$--GRU and Rad$^{phys}$--GRU models exhibit hidden-state dynamics that change primarily in the alarm-positive regime, consistent with their corresponding feature-activation patterns (Figs.~\ref{fig:hidden}B and C). For CNN$^{cl}$--GRU at small $E$, $v_t^h$ displays a ReLU-like response across the alarm boundary, with appreciable changes emerging mainly after $E=D_c$. At large $E$, $v_t^h$ shows little variation across the alarm boundary, while appreciable changes in latent dynamics mainly emerge in the alarm-positive regime. The Rad$^{phys}$--GRU exhibits only modest variations in $v_t^h$ across the alarm boundary, making the transition difficult to identify from the hidden-state dynamics alone.

Overall, these observations suggest that the hidden state of the CNN--GRU provides a low-dimensional surrogate for the evolving morphological state, capturing the progressive destabilization of compact growth prior to the emergence of a fully dendritic morphology. From this perspective, the CNN--GRU can be viewed as performing an implicit coarse-graining of the high-dimensional growth dynamics, mapping microscopic morphological fluctuations onto a low-dimensional latent trajectory. This interpretation further supports the importance of simultaneous spatiotemporal learning for early-warning prediction of the CDT.

At the same time, the emergence of meaningful latent-space dynamics does not rely uniquely on the presence of a recurrent unit. The latent-space speed of the CNN--TCN (Figs.~\ref{fig:hidden}D and H) also exhibits systematic changes near the alarm-boundary, qualitatively similar--though less pronounced--to the slowing-down behavior observed for the GRU-based models. This suggests that the key ingredient is the end-to-end spatiotemporal learning framework itself, rather than a specific choice of temporal integration mechanism.

\subsection{Transferability to other reaction-rate conditions: Limited yet systematic}

Finally, we evaluate the CNN--GRU model trained at $\log_{10}k=-1.69$ under two additional reaction-rate conditions, $\log_{10}k=-1.90$ and $-2.12$, which correspond to slower reaction kinetics and therefore a later onset of the CDT during growth. As shown in Fig.~\ref{fig:cdt_morph}B–C, CDTs arising under different reaction--diffusion conditions can be collapsed onto a single renormalized representation based on statistical-mechanical scaling relations. Such collapse might suggest that a model trained under one condition could generalize across reaction and diffusion rates.

In contrast to this physics-motivated expectation, the CNN–GRU model pre-trained at $\log_{10} k=-1.69$ exhibits limited transferability to other reaction-rate conditions, albeit in a systematic manner (see Fig.~\ref{si:fig:var_E_wo_ft} in the Appendix). As the reaction rate departs from the training condition toward $\log_{10} k=-1.90$ and $-2.12$, the F1-score performance degrades substantially.

For $\log_{10} k=-1.90$, the F1-negative score in the alarm-negative regime remains relatively high and decreases systematically with increasing $E$. In contrast, performance in the alarm-positive regime deteriorates more strongly. At $\log_{10} k=-2.12$, forecasting performance degrades severely in both alarm-negative and alarm-positive regimes.

This systematic, $k$-dependent transferability is further reflected in the latent-state dynamics. For $\log_{10} k=-1.90$, the latent-space speed of the hidden state exhibits a slowing-down behavior near the alarm boundary similar to that of the pre-trained case, but lacks sufficient resolution to reliably pinpoint the boundary at $E=D_c$. For $\log_{10} k=-2.12$, the latent dynamics no longer show a meaningful reorganization around the alarm boundary, consistent with the strongly degraded predictive performance.

Although fine-tuning the pre-trained model at $\log_{10}k=-1.69$ restores strong predictive performance in both cases (see Fig.~\ref{si:fig:var_E_w_ft} in the Appendix), we find no evidence for accelerated learning during fine-tuning. These results suggest that, despite the statistical-mechanical collapse of the CDT under reaction-rate rescaling, the learned latent representation does not inherit this invariance. Consequently, reliable forecasting appears to require explicit retraining for each reaction-rate condition. A more detailed analysis of this limitation is left for future work.

\section{Conclusions and Future Directions}

In this work, we investigated compact-to-dendritic transitions (CDTs) in a controlled particle-based electrodeposition model and formulated an explicit early-warning prediction problem based on trajectory-resolved transition points. By combining physics-based growth simulations with image-based spatiotemporal learning, we demonstrated that early-warning of dendritic instability is intrinsically a spatiotemporal problem: neither static morphological descriptors nor temporal learning built on predefined features alone provide reliable precursors. Instead, end-to-end learning of joint spatial and temporal representations (CNN--GRU and CNN--TCN) enables robust anticipation of the transition over extended prediction horizons. 

Analysis of learned features and hidden-state dynamics further indicates that the CNN--GRU constructs a low-dimensional surrogate of the evolving morphological state, consistent with an implicit coarse-graining of nonequilibrium growth dynamics. We further showed that this learned spatiotemporal representation exhibits limited but systematic transferability across reaction-rate conditions, with predictive performance degrading as the inference condition departs from the training condition, consistent with corresponding changes in latent-state dynamics.

Several directions merit future investigation. A natural extension of the present two-dimensional framework is to three-dimensional deposition, where morphological complexity and heterogeneity are further amplified and early detection of incipient instabilities is even more challenging. Hybrid approaches that combine learned spatiotemporal representations with physics-informed features may improve interpretability while retaining high early-warning performance. Extending the framework to experimental imaging data, including noisy or partially observed interfaces, will be essential for practical deployment.

Physically informed early-warning and control strategies may ultimately provide an important component for electrochemical energy-storage systems operating near their stability limits. Achieving this goal requires the ability to characterize and interpret latent-state dynamics, opening a pathway toward dynamic control of nonequilibrium growth processes. If the latent state indeed tracks the evolving morphological state, it could serve as a compact surrogate for feedback and closed-loop control, enabling charging or deposition protocols to be adaptively modulated to steer the system away from catastrophic dendritic growth~\cite{nielsen_sharp-interface_2015, tikekar_stabilizing_2016, choudhury_confining_2018}.

\appendix

\renewcommand{\thefigure}{A\arabic{figure}}
\renewcommand{\thetable}{A\arabic{table}}
\setcounter{figure}{0}
\setcounter{table}{0}

\section{Particle-based model for electrodeposition in two dimensions}\label{sec:appendix:particle}

Battery charging is a complex process involving multiple coupled molecular phenomena, including bulk and interfacial ion diffusion and electroreduction~\cite{macfarlane_lithium-doped_1999, macfarlane_lithium-doped_1999, jacobsonContinuumLimitDendritic2025b, nielsen_sharp-interface_2015}. In this work, we intentionally simplify this process by modeling electrodeposition as particle-based aggregation in 2D~\cite{jacobsonContinuumLimitDendritic2025b}, represented as a sequence of stochastic aggregation events. Each event consists of two elementary processes: Brownian diffusion of a reactive particle in the electrolyte domain and a probabilistic deposition reaction upon contact with the growing aggregate. The latter represents the electrochemical reduction at the battery interface (e.g., $\mathrm{Li}^+ +~e^- \rightarrow \mathrm{Li}^0$ in the case of a lithium metal battery). In this work, we adopt a publicly available implementation of the model and the parameters provided in Ref.~\onlinecite{DanGithub}.

\textbf{Brownian diffusion.}
Following Ref.~\onlinecite{jacobsonContinuumLimitDendritic2025b}, an algorithmic treatment is employed to enhance computational efficiency while preserving the correct diffusive statistics. Reactive particles are launched from a bounding surface enclosing the existing aggregate and undergo Brownian motion governed by a diffusion coefficient $D$. When a particle is far from the aggregate, its motion is accelerated using a first-passage jump scheme. Specifically, if the minimum distance $H$ between the particle and the aggregate exceeds a cutoff distance $\sigma_{\mathrm{cut}}$, the particle is displaced by a uniformly distributed jump of radius $d_{\mathrm{jump}} = H - \varepsilon$, where $\varepsilon$ is a small numerical buffer. Once the particle enters the near-field region ($H \le \sigma_{\mathrm{cut}}$), its motion is resolved explicitly via Gaussian-distributed displacements,
\begin{equation}
  \boldsymbol{r}(t+\Delta t) = \boldsymbol{r}(t) + \mathcal{N}(0, 2D\Delta t\,\mathbf{I}),
\end{equation}
where $\Delta t$ denotes the simulation time step.

\textbf{Deposition reaction.}
If a collision with the aggregate is detected, a deposition reaction occurs with probability
\begin{equation}
  P = k \sqrt{\frac{\pi \Delta t}{D}},
\end{equation}
where $k$ is the reaction rate constant. With probability $P$, the particle irreversibly sticks to the aggregate and becomes part of the growing deposit; otherwise, it undergoes elastic reflection and continues diffusing \cite{jacobsonContinuumLimitDendritic2025b}. This probabilistic sticking rule allows the model to continuously interpolate between reaction-limited growth ($P \ll 1$), which produces compact morphologies, and diffusion-limited growth ($P \to 1$), which leads to dendritic structures. In this work, we vary only the reaction rate constant $k$ while keeping all other parameters fixed, enabling a systematic investigation of morphological instabilities associated with the compact-to-dendritic transition in electrodeposition \cite{jacobsonContinuumLimitDendritic2025b}.

\begin{table*}[t]\centering\label{tab:params_dla}
  \begin{tabular}{llll}
    \hline
    \textbf{Parameter} & \textbf{Symbol} & \textbf{Value} & \textbf{Description} \\
    \hline
    Reaction rate constant & $k$ & $10^{-1.69}, 10^{-1.9}, 10^{-2.12}$ & Controls sticking probability \\
    Diffusion coefficient & $D$ & $1.0$ & Particle diffusion coefficient \\
    Particle radius & $a$ & $1.0$ & Effective particle radius \\
    Time step & $\Delta t$ & $10^{-3}$ & Brownian integration step \\
    Cutoff distance & $\sigma_{\mathrm{cut}}$ & $10~a$ & Far-/near-field threshold \\
    Numerical buffer & $\varepsilon$ & $10^{-3}~a$ & Jump-distance stabilizer \\
    System size & $L$ & $1024~a$ & Simulation box size \\
    Total particle number & $N_{\mathrm{total}}$ & $10^{6}$ & Particles per realization \\
    \hline
  \end{tabular}
  \caption{Simulation parameters used in the particle-based electrodeposition model.}
\end{table*}

\section{2D growth images using grid representation}\label{sec:appendix:images_pre}

The particle coordinates are discretized onto a fixed square grid to obtain binary occupancy images, where occupied and unoccupied pixels represent the presence or absence of deposited particles. The spatial bounds are determined using the complete particle set to ensure consistent spatial scaling across all time points. The projected coordinates are then mapped onto a $H \times W$ grid, following a global coordinate normalization. In this work, we only  use square grids with $H=W$.

\begin{equation}
  \text{coord}_{\mathrm{norm}} = 
  \frac{\text{coord} - \text{min}_{\mathrm{bound}}}{\text{range}_{\mathrm{bound}}},
\end{equation}

which ensures that all images share a common spatial reference frame. A reference configuration containing $10^6$ particles is used to define the global normalization and scaling parameters, which are applied consistently across all growth stages.

A physical length scale is further defined as
\begin{equation}\label{eq:lengconv}
  \text{scale} = \frac{\min(x_{\mathrm{range}}, y_{\mathrm{range}})}{\text{grid size}},
\end{equation}
allowing conversion between pixel units and physical distances in subsequent quantitative analyses.

\begin{figure*}[htbp!]\centering
  \includegraphics[width=5.5in]{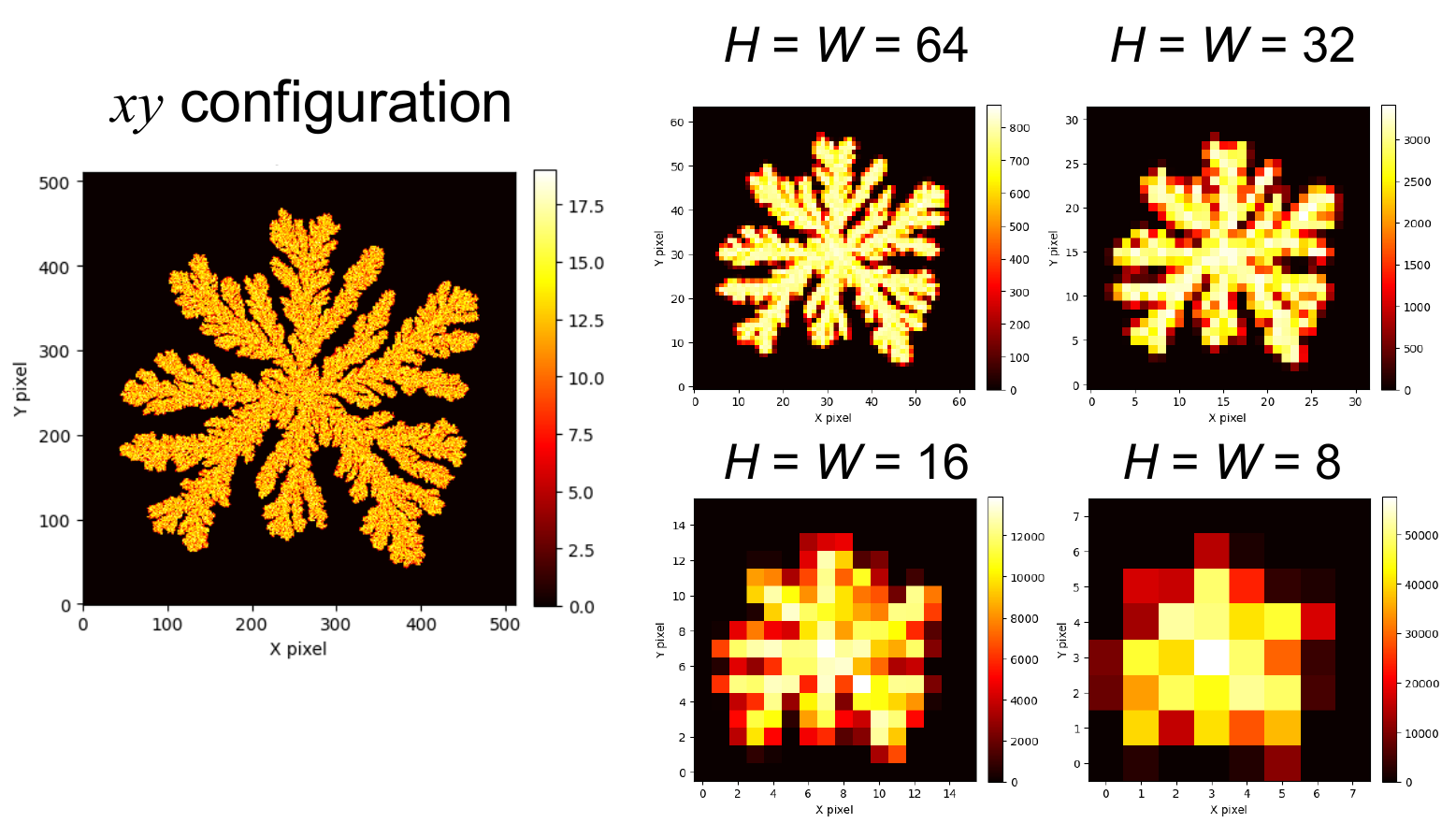}
\caption{Grid-based representation of the growth images. The colorbar indicates the particle density in each pixel for different spatial resolutions $H\times W$.} \label{si:fig:grid}
\end{figure*}

This grid-based representation $G$ provides a standardized and scale-consistent description of evolving deposition morphologies, enabling direct quantitative comparison of spatial patterns and ensuring compatibility with both physics-based diagnostics and data-driven learning methods (Fig.~\ref{si:fig:grid}). The same image construction, normalization, and spatial scaling procedures are applied consistently throughout this work.

\section{Physics-based characterization}\label{sec:appendix:images_phys}

We quantify aggregate morphologies using a set of interrelated, physically interpretable descriptors, including the fractal dimension, interfacial length, circularity, aspect ratio, and convex hull ratio. Except for the fractal dimension, all physical quantities are computed from the binary grid images using \texttt{OpenCV} (version~4.11.0.86).

\textbf{Fractal dimension, $d_f$:}
The fractal dimension characterizes the global scaling properties of the aggregate through the relation in Eq.~\ref{eq:fractal} in the main text. In practice, the derivative to compute $d_f$ in Eq.~\ref{eq:df} in the main text is evaluated using a sliding-window approach centered at each particle count, incorporating data from $\pm 4$ neighboring sampling steps. The step size is chosen as $\max(5{,}000, N/8)$ to balance numerical stability and computational efficiency. To suppress unphysical values arising statistical noise, the estimated values are restricted to $1.5 \le d_f \le 2.2$, corresponding to physically realistic ranges for two-dimensional diffusion-limited aggregation.

\textbf{Interfacial length, $L$:}
The interfacial length is defined as the total perimeter of the aggregate and provides a measure of surface roughness and branching complexity. The perimeter is extracted from binary occupancy images using Canny edge detection, as implemented in \texttt{OpenCV}, with parameters $\sigma = 2.0$, $\text{low\_threshold} = 50$, and $\text{high\_threshold} = 150$. The pixel-based perimeter is converted to physical units using the spatial scale factor defined in Eq.~\ref{eq:lengconv}.

\textbf{Circularity, $C$:}
Circularity quantifies the compactness of the aggregate shape and is defined as
\begin{equation}
  C = \frac{4\pi A}{P^2},
\end{equation}
where $A$ is the area of the aggregate and $P$ is its perimeter. A perfect circle yields $C=1$, whereas irregular or fractal structures exhibit smaller values. Circularity is computed from the contours of the binary grid representation and provides a simple geometric measure of morphological irregularity.

\textbf{Aspect ratio, AR:}
The aspect ratio characterizes directional anisotropy in aggregate growth. Principal component analysis (PCA) is applied to the spatial distribution of occupied grid points to extract the major and minor principal axes. The aspect ratio is defined as the ratio of the minor-axis length to the major-axis length. Values significantly smaller than unity indicate elongated or directionally biased growth, whereas values close to unity correspond to isotropic morphologies.

\textbf{Convex hull ratio, CHR:}
The convex hull ratio quantifies branching and porosity of the aggregate and is defined as
\begin{equation}
  CHR = \frac{A}{A_{\mathrm{CH}}},
\end{equation}
where $A_{\mathrm{CH}}$ denotes the area of the convex hull. Lower values of CHR indicate more branched and fractal morphologies, while values approaching unity correspond to compact, space-filling growth.




\section{Computation of normalized activation and feature importance}
\label{sec:appendix:activ_norm}

Each input image is processed by a CNN with $S_F=16$ channels, and each channel output is reduced to a scalar activation $e_c$ through global average pooling over spatial dimensions. To facilitate comparisons across samples, we normalize activations channel-wise over the full dataset as
\begin{equation}
\tilde{e}_c^{(n)} =
\frac{e_c^{(n)} - \min_k e_c^{(k)}}
{\max_k e_c^{(k)} - \min_k e_c^{(k)}} \in [0,1],
\end{equation}
where $n$ labels samples and $k$ runs over all samples.

The normalized activations are examined as a function of the control parameter $\log_{10}(k R_g/D)$, which is discretized into 50 uniform bins. Within each bin $b$ and for each channel $c$, we compute the mean $\mu_{c,b}$ and standard deviation $\sigma_{c,b}$ of $\tilde{e}_c$.

Channel relevance is quantified using gradient-based saliency. Specifically, we define an importance score
\begin{equation}
I_c = \mathbb{E}\!\left[\left| \frac{\partial \hat{y}}{\partial \mathbf{F}_c} \right|\right],
\end{equation}
where $\hat{y}$ denotes the model output and $\mathbf{F}_c$ the feature map of channel $c$. Channels are ranked according to $I_c$, and the five most important channels are chosen and visualized.

\begin{figure}[htbp!]\centering
  \includegraphics[width=3.3in]{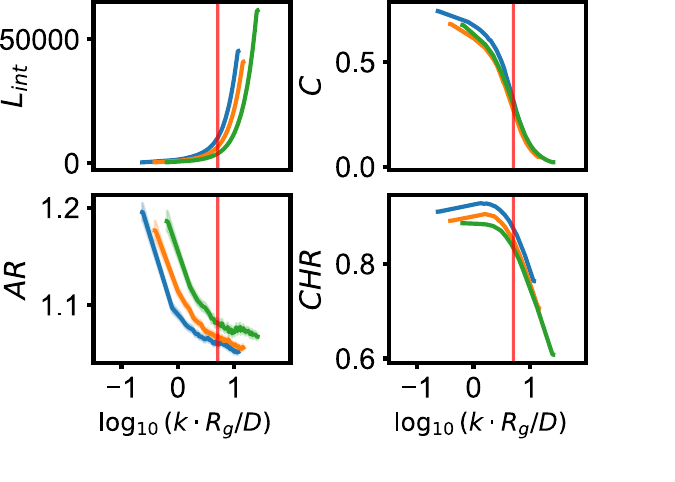}
  \caption{Physical characterizations, including interfacial length ($L_{int}$), circularity ($C$), aspect ratio (AR), and convex hull ratio (CHR). As a visual reference, red vertical lines indicate the CDT at $R_g = R_c$, adopting the ensemble-averaged value of $R_c$ reported in Ref.~\onlinecite{jacobsonContinuumLimitDendritic2025b}.} \label{si:fig:phys}
\end{figure}

\begin{figure*}[htbp!]\centering
  \includegraphics[width=5.5in]{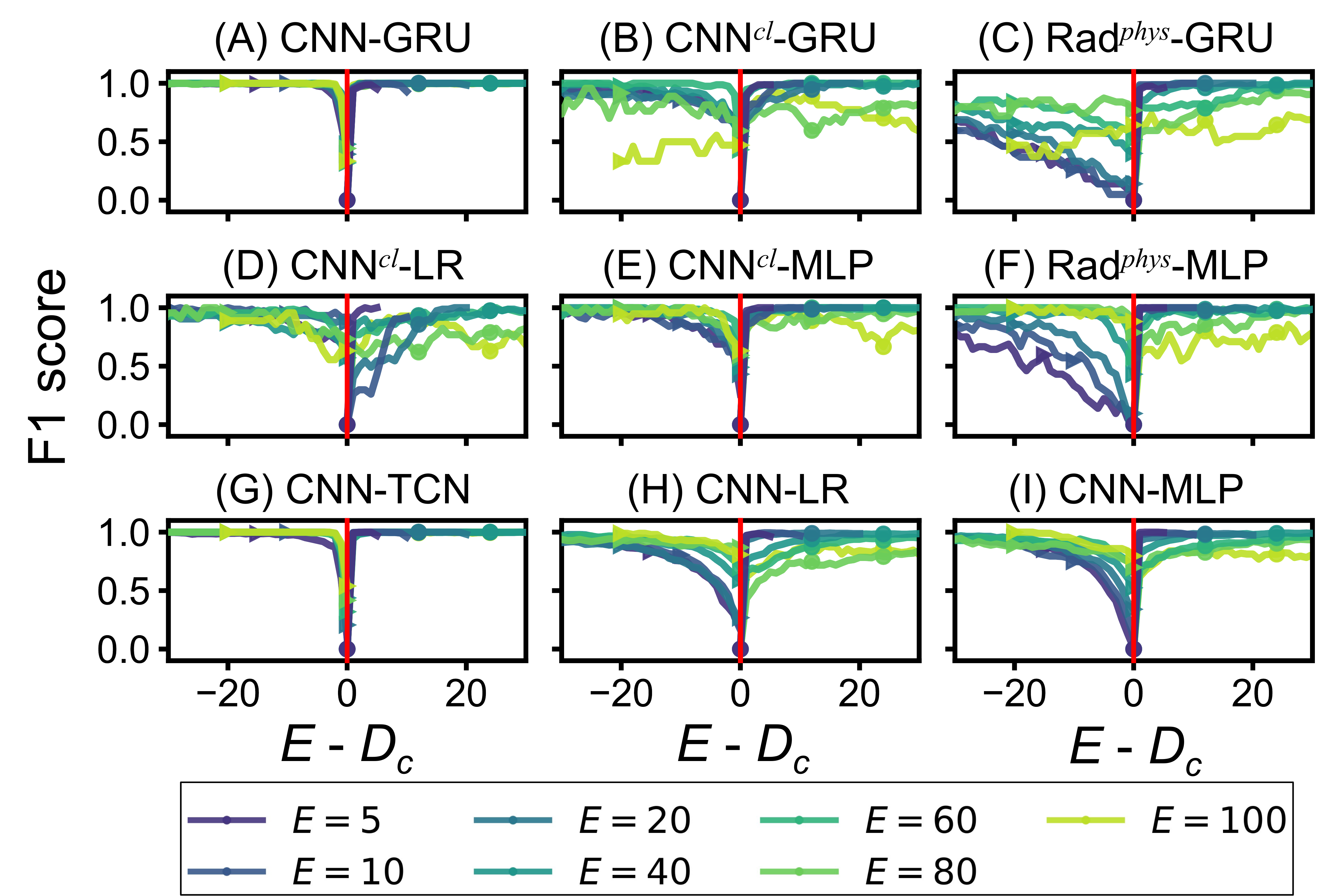}
  \caption{F1-score performance of early-warning models as a function of $E - D_c$ at $L_s = 5$. Circle and triangle markers denote the F1-positive and F1-negative scores, respectively. The panels show magnified views of the corresponding results in Fig.~\ref{fig:main_results} of the main text.}
  \label{si:fig:f1_zoom}
\end{figure*}

\begin{figure*}[htbp!]\centering
  \includegraphics[width=5.5in]{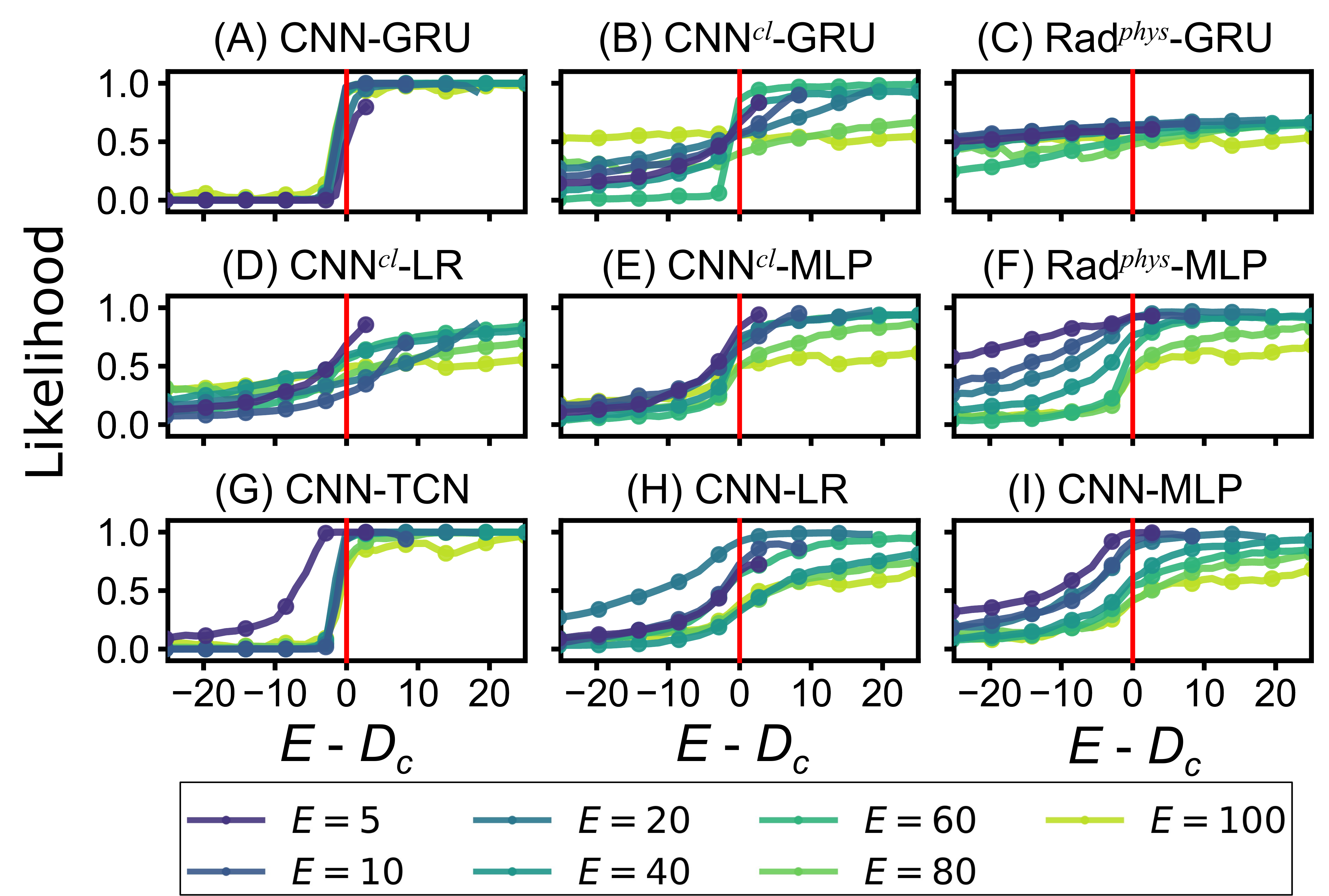}
\caption{Likelihood evaluation of early-warning models as a function of $E - D_c$ at $L_s = 5$. The panels show the corresponding results in Fig.~\ref{fig:main_results} of the main text.}
  \label{si:fig:likelihood}
\end{figure*}

\begin{figure*}[htbp!]\centering
  \includegraphics[width=5.5in]{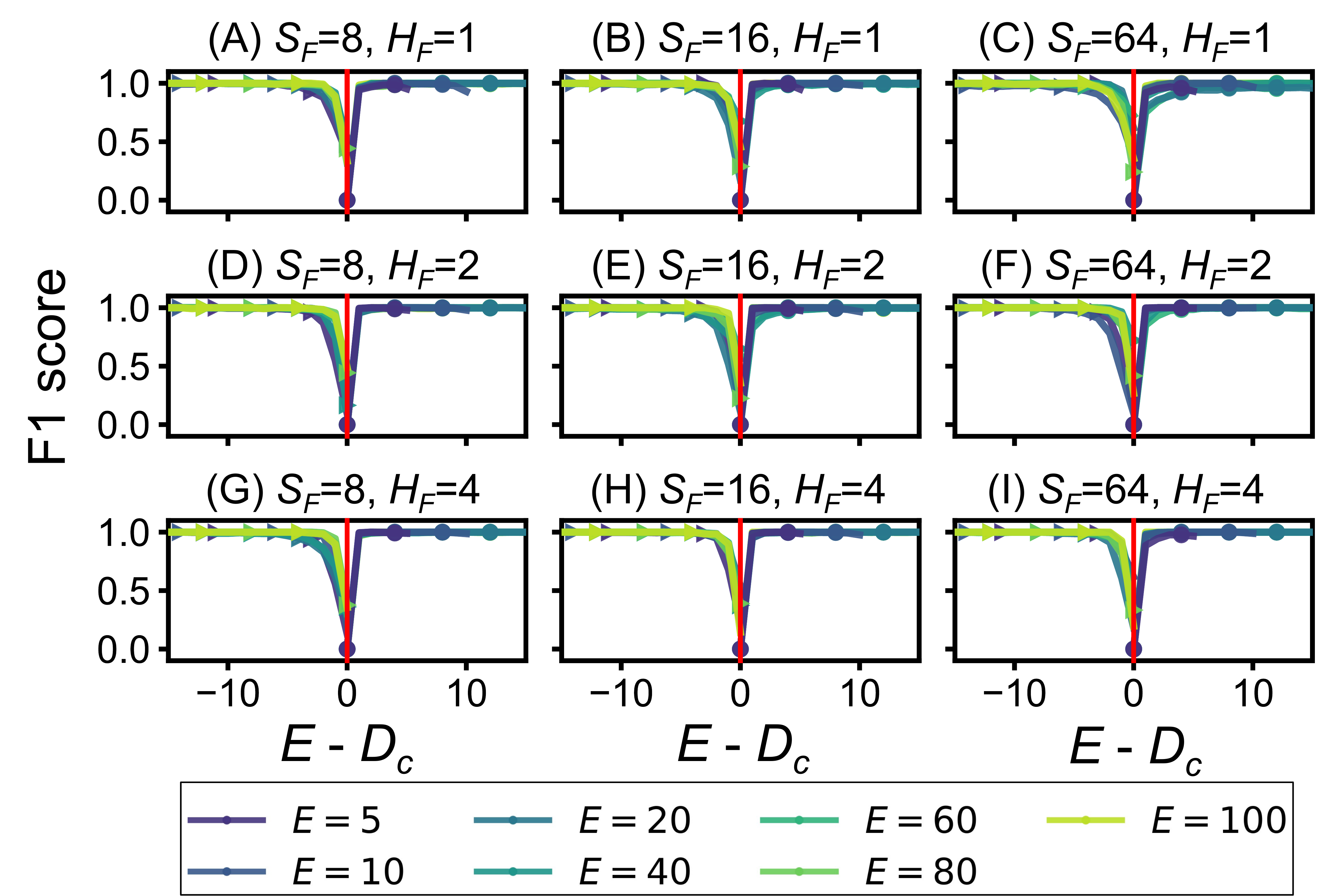}
\caption{F1-score performance of early-warning models as a function of $E - D_c$ at $L_s = 5$ with varying spatial information: CNN feature dimension $S_F$ and global spatial pooling $H_F\times W_F$. Circle and triangle markers denote the F1-positive and F1-negative scores, respectively. Here, $H_F=W_F$.}
  \label{si:fig:f1_var}
\end{figure*}

\begin{figure*}[htbp!]\centering
  \includegraphics[width=5.5in]{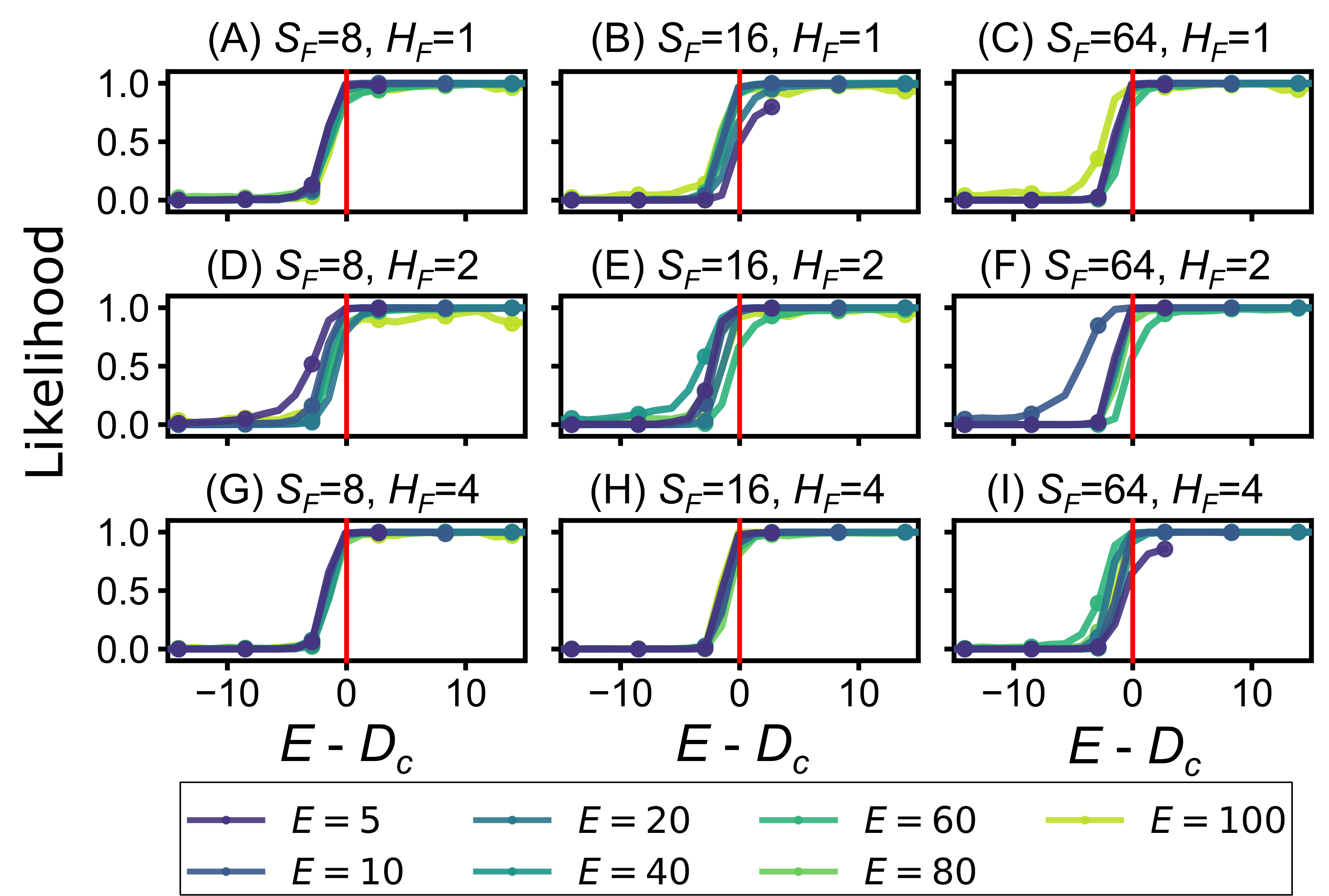}
\caption{Likelihood evaluation of early-warning models as a function of $E - D_c$ at $L_s = 5$ with varying spatial information: CNN feature dimension $S_F$ and global spatial pooling $H_F\times W_F$.  Here, $H_F=W_F$. The panels show the corresponding results in Fig.~\ref{si:fig:f1_var}.}  \label{si:fig:likelihood_var}
\end{figure*}

\begin{figure*}[htbp!]\centering
  \includegraphics[width=6in]{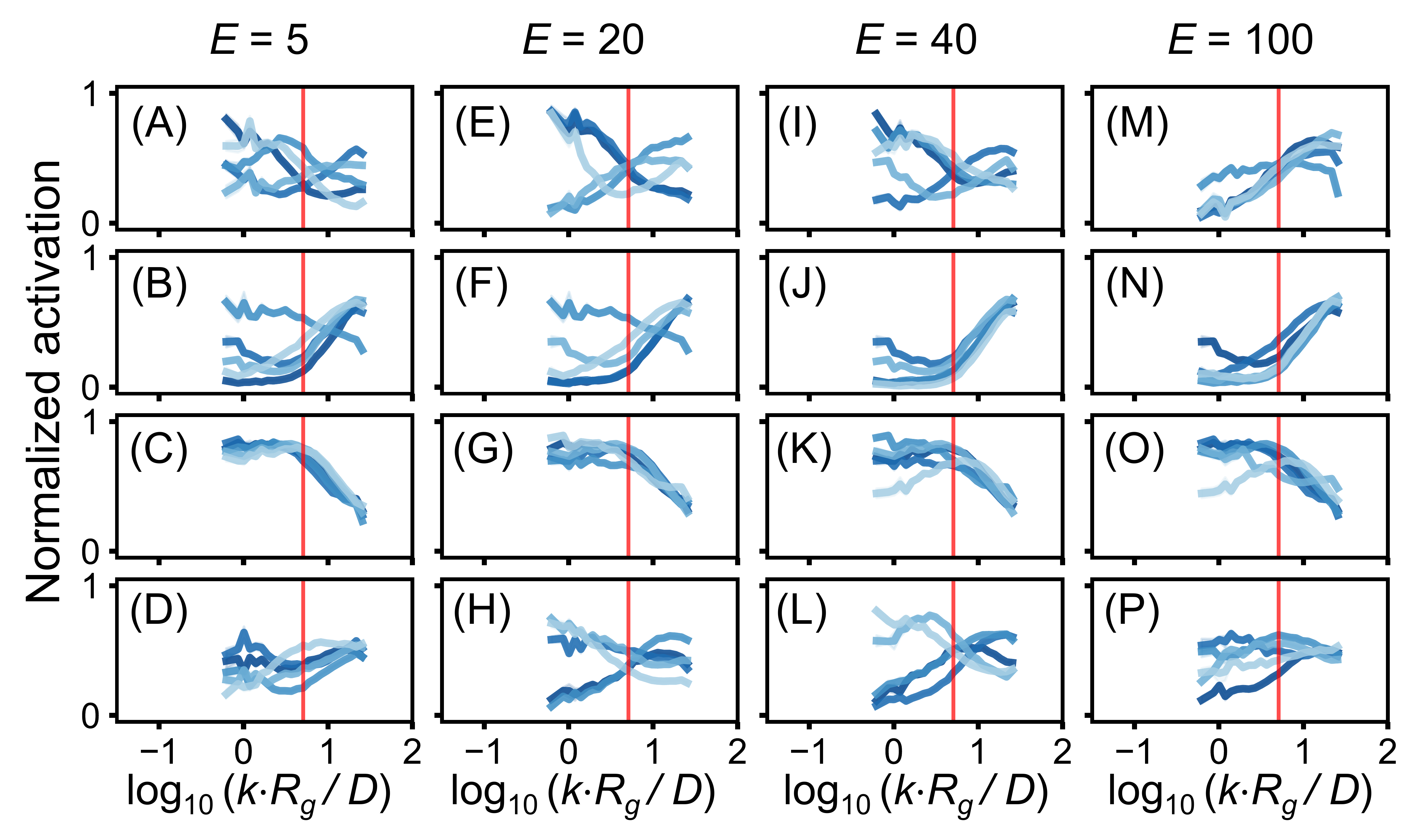}
    \caption{Normalized activation of spatial features $e_t$ for different prediction horizons $E$ during growth at $\log_{10} k=-1.69$. 
(A,E,I,M) CNN--GRU, (B,F,J,N) CNN$^{cl}$, (C,G,K,O) Rad$^{phys}$, and (D,H,L,P) CNN--TCN. 
Results for $E=5$ correspond to Fig.~\ref{fig:embedding} in the main text. 
As a visual reference, red vertical lines indicate the CDT at $R_g = R_c$, adopting the ensemble-averaged value of $R_c$ reported in Ref.~\onlinecite{jacobsonContinuumLimitDendritic2025b}. 
Bluer colors indicate features with higher importance scores.}
  \label{si:fig:activation_grid}
\end{figure*}

\begin{figure*}[htbp!]\centering
  \includegraphics[width=6in]{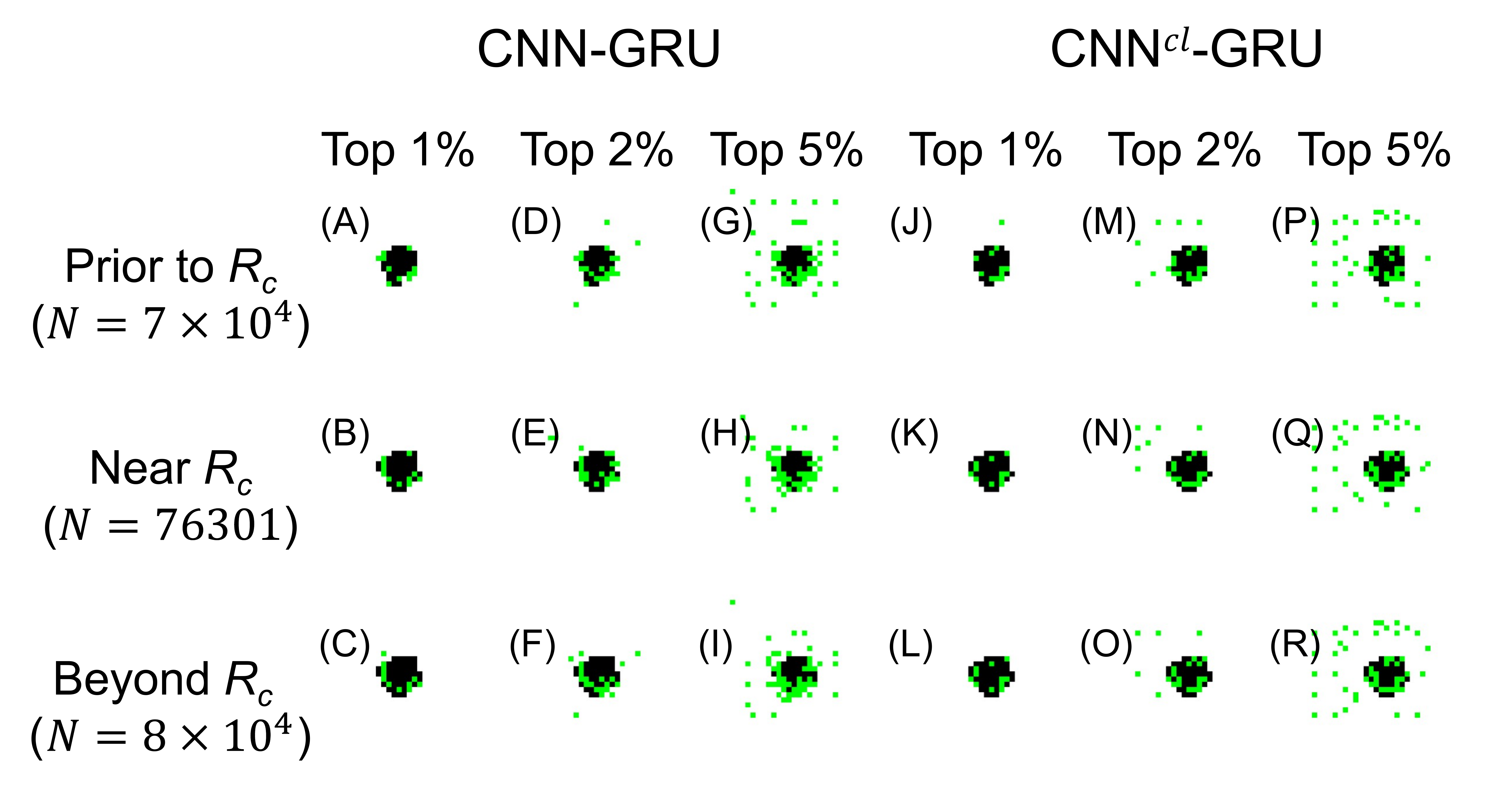}
  \caption{Saliency analysis of CNN features using guided backpropagation. Deposit pixels are shown in black, while pixels with high saliency are highlighted in bright green.}
  \label{si:fig:gbp}
\end{figure*}

\begin{figure*}[htbp!]\centering
  \includegraphics[width=6in]{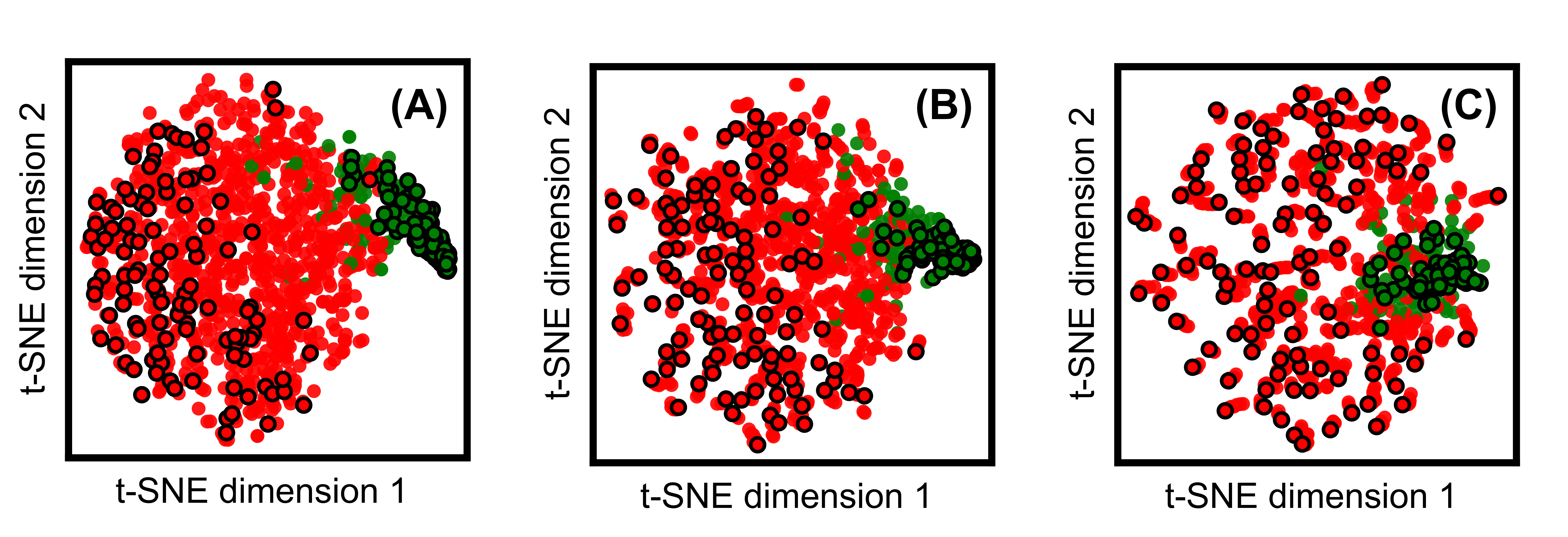}
\caption{Two-dimensional t-SNE embeddings of 100 growth trajectories at $\log_{10} k=-1.69$ constructed from $e_t$ of the CNN--GRU model. The panels correspond to Fig.~\ref{fig:main_results}A in the main text and show different global spatial pooling resolutions: (A) $1\times1$, (B) $2\times2$, and (C) $4\times4$. Markers with black edges denote the starting and end points of each trajectory, and marker color indicates the morphological state: green for compact growth and red for dendritic growth.}
  \label{si:fig:tsne_all}
\end{figure*}

\begin{figure*}[htbp!]\centering
  \includegraphics[width=5.5in]{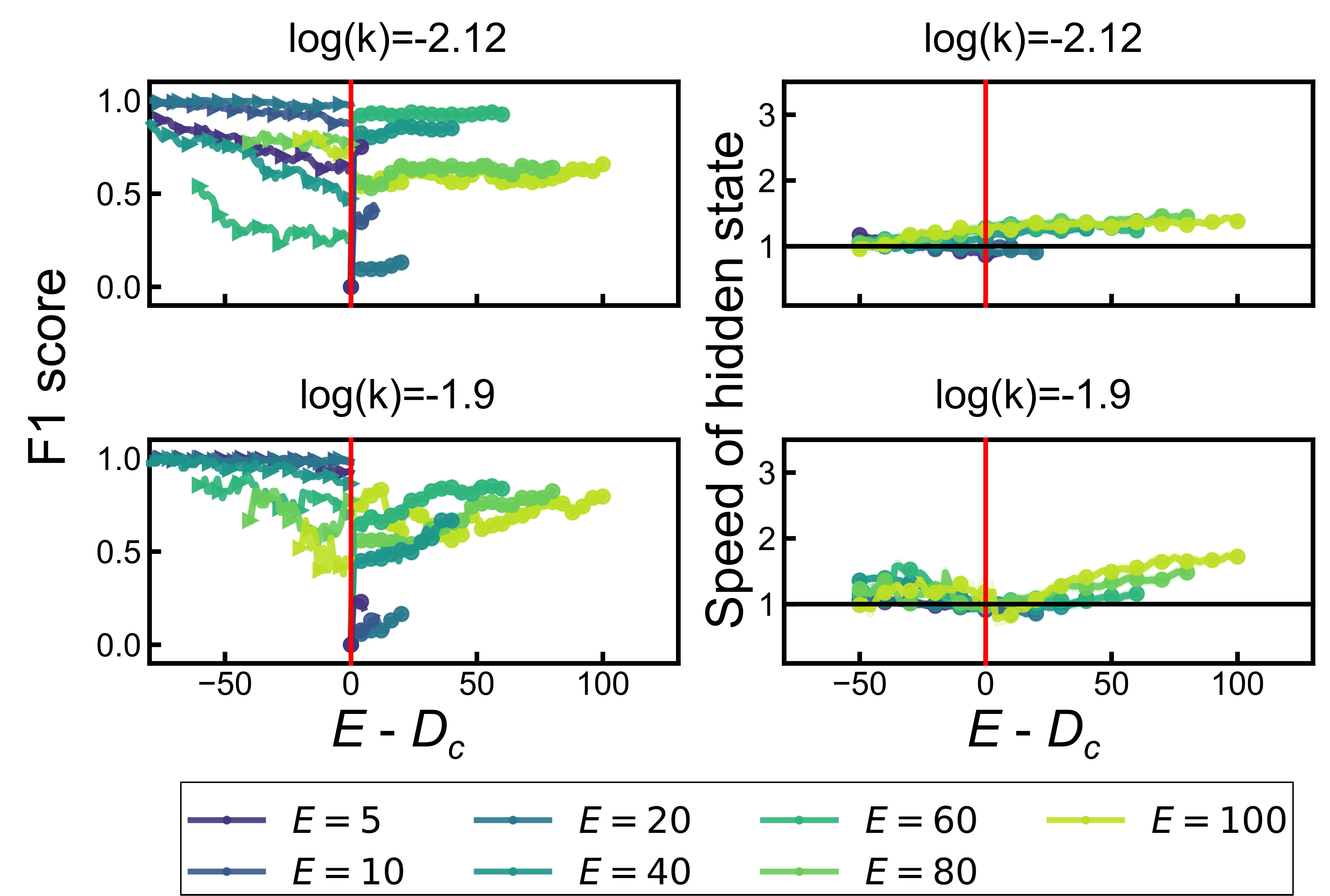}
\caption{Transfer inference using a CNN--GRU model without fine-tuning. The model is pre-trained at $\log_{10} k=-1.69$, corresponding to Fig.~\ref{fig:main_results} in the main text.} \label{si:fig:var_E_wo_ft}
\end{figure*}

\begin{figure*}[htbp!]\centering
  \includegraphics[width=5.5in]{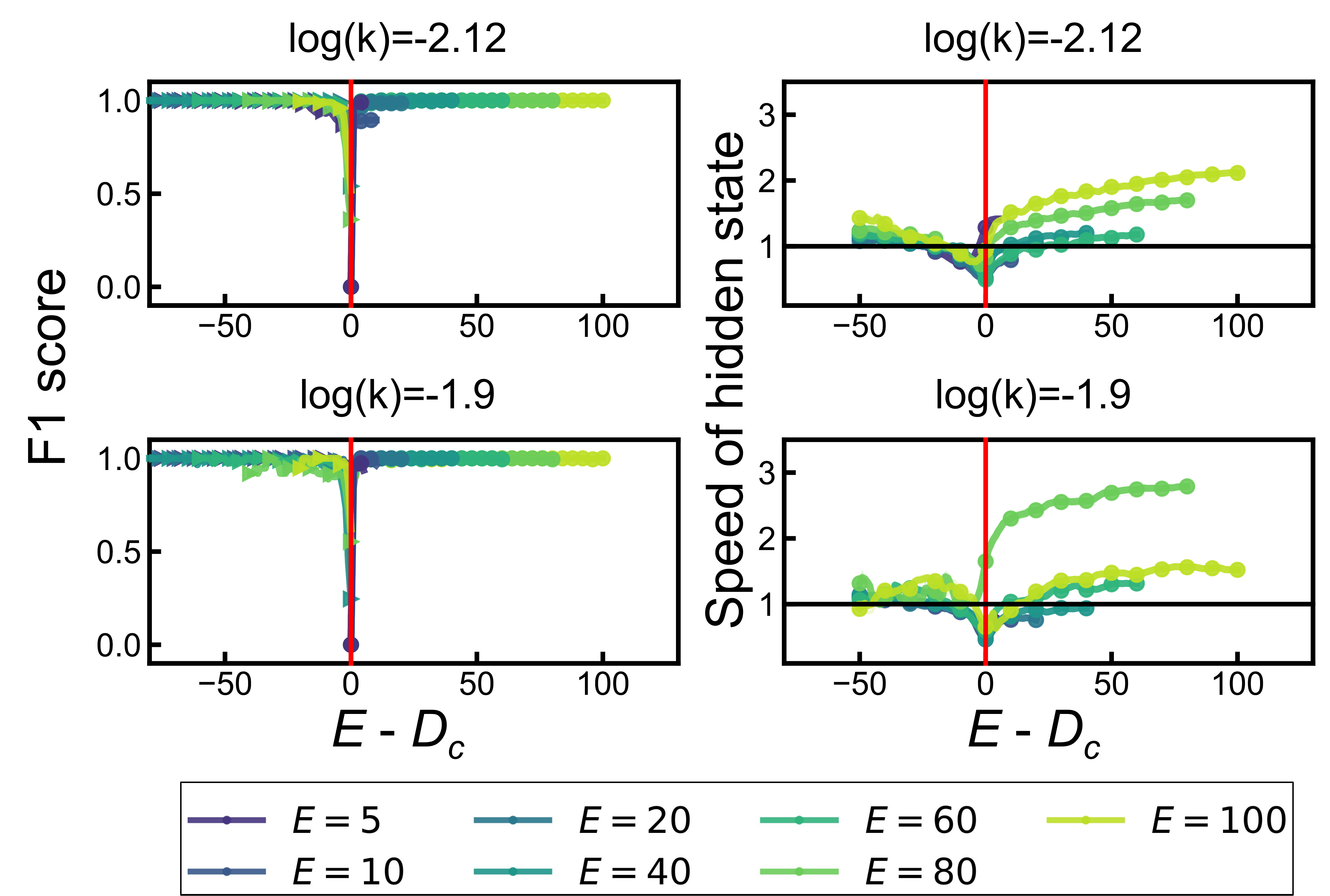}
\caption{Transfer inference using a CNN--GRU model with fine-tuning. The model is pre-trained at $\log_{10} k=-1.69$, corresponding to Fig.~\ref{fig:main_results} in the main text.}  \label{si:fig:var_E_w_ft}
\end{figure*}

\begin{figure*}[htbp!]\centering
  \includegraphics[width=5.5in]{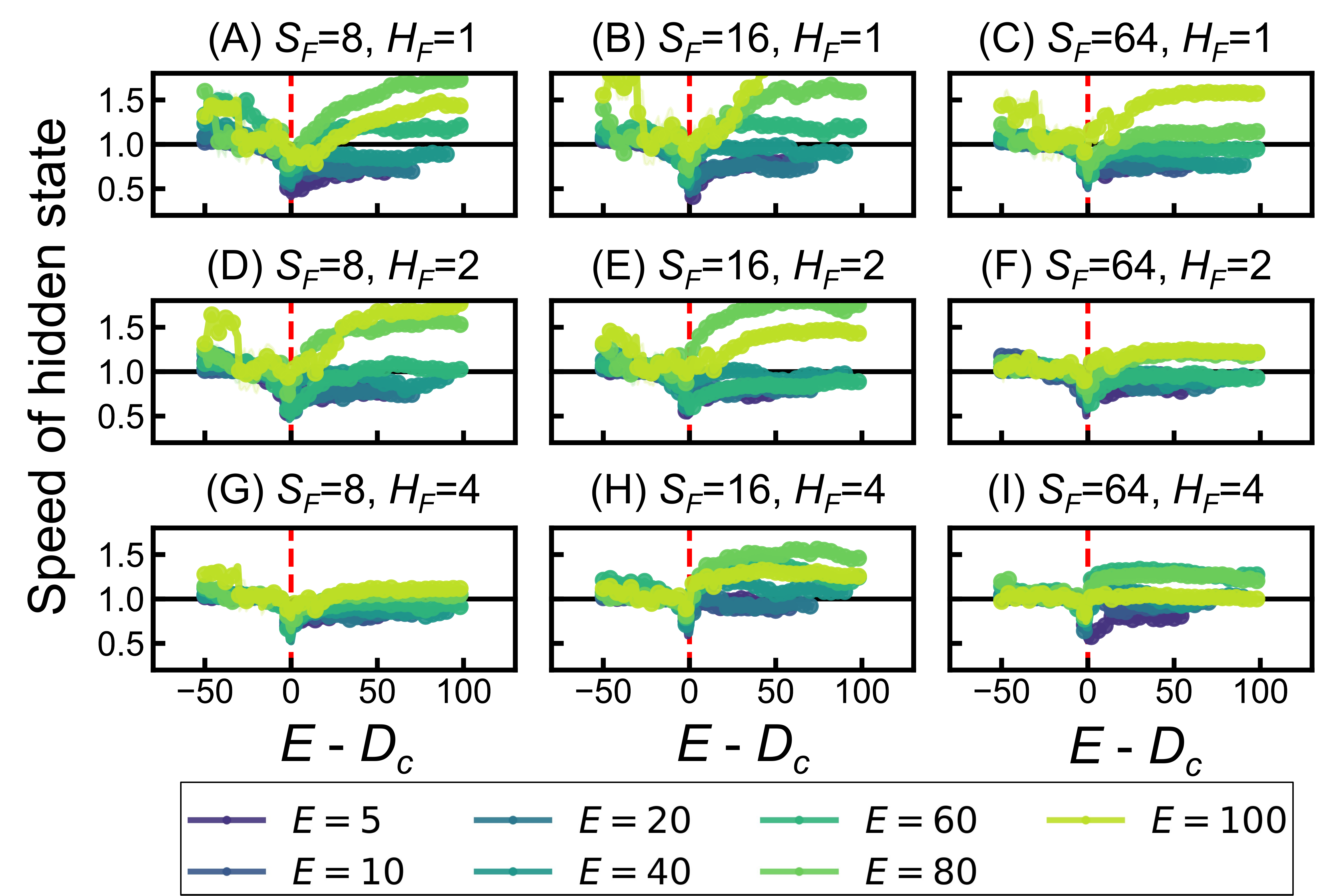}
\caption{Latent-space speed of hidden hidden state during growth with varying spatial information: CNN feature dimension $S_F$ and global spatial pooling $H_F\times W_F$.  Here, $H_F=W_F$. The panels show the corresponding results in Figs.~\ref{si:fig:f1_var} and~\ref{si:fig:likelihood_var}.} \label{si:fig:hidden_var}
\end{figure*}


\bibliography{main_cleaned}

\end{document}